\documentclass[10pt,a4paper,twocolumn]{article}

\usepackage{cite}
\usepackage{amsmath,amssymb,amsfonts}
\usepackage{algorithmic}
\usepackage{graphicx}
\usepackage{textcomp}

\usepackage{amsthm}
\usepackage{pifont}
\usepackage{mathtools}
\usepackage{float}
\usepackage{caption} 
\usepackage{subcaption}
\usepackage{tabularx}
\usepackage{listings}
\usepackage{float}

\usepackage{booktabs}
\usepackage{quantikz}
\usepackage{tikz}

\usetikzlibrary{shapes.geometric, arrows}

\tikzstyle{startstop} = [rectangle, rounded corners, minimum width=3cm, minimum height=1cm,text centered, draw=black, fill=red!30]
\tikzstyle{process} = [rectangle, minimum width=3cm, minimum height=1cm, text centered, draw=black, fill=orange!30]
\tikzstyle{decision} = [diamond, minimum width=3cm, minimum height=1cm, text centered, draw=black, fill=green!30]
\tikzstyle{arrow} = [thick,->,>=stealth]

\def\exeins{\ensuremath{\mathsf{exe_{ins}}}}
\def\exe{\ensuremath{\mathsf{exe}}}

\usepackage{diagbox}
\usepackage{lscape}
\usepackage{bm}

\usepackage{hyperref}

\usepackage{threeparttable}
\usepackage{multirow}
\usepackage[export]{adjustbox}

\newcommand{\numParam}{\ensuremath{\mathsf{numParam}}}
\newcommand{\numQubits}{\ensuremath{\mathsf{numQbts}}}
\newcommand{\MutableGateSet}{\ensuremath{\mathsf{MutSet}}}
\newcommand{\name}{\ensuremath{\mathsf{name}}}
\newcommand{\Exc}{\ensuremath{\mathsf{Exc}}}
\newcommand{\Op}{\ensuremath{\mathsf{Op}}}
\newcommand{\Del}{\ensuremath{\mathsf{Del}}}
\newcommand{\Name}{\ensuremath{\mathsf{Name}}}
\newcommand{\Attr}{\ensuremath{\mathsf{Attr}}}
\newcommand{\Ins}{\ensuremath{\mathsf{Ins}}}
\newcommand{\bIns}{\ensuremath{\bm{\mathsf{Ins}}}}

\newcommand{\allins}{\ensuremath{\mathsf{All\_Ins\_Data}}}
\newcommand{\reduction}{\ensuremath{\mathsf{Reduction}}}
\newcommand{\breduction}{\ensuremath{\bm{\mathsf{Reduction}}}}
\newcommand{\cmark}{\ding{51}}
\newcommand{\xmark}{\ding{55}}

\theoremstyle{definition}
\newtheorem{definition}{Definition}[section]
\newtheorem{proposition}{Proposition}[section]
\newtheorem{example}{Example}[section]

\newfloat{lstfloat}{htbp}{lop}
\floatname{lstfloat}{Listing}

\def\BibTeX{{\rm B\kern-.05em{\sc i\kern-.025em b}\kern-.08em
    T\kern-.1667em\lower.7ex\hbox{E}\kern-.125emX}}

\title{QCRMut: Quantum Circuit Random Mutant generator tool}

\begin{document}


\author{Sinhu\'{e} Garc\'{i}a Gil$^{1}$, Luis Llana$^{1}$, and Jos\'{e} Ignacio Requeno Jarabo$^{1}$ \\
\{sinugarc, llana, jrequeno\}@ucm.es\\ \\
$^{1}$\textit{Facultad de Inform\'{a}tica, Universidad Complutense de Madrid} \\ 
\textit{Calle Jos\'{e} Garc\'{i}a Santesmases, 9, 28040, Madrid, Spain}}

\maketitle

\begin{abstract}

As quantum computing moves towards practical deployment, ensuring the reliability of quantum software becomes increasingly important. Mutation testing is a promising technique in this context; however, existing exhaustive mutation generators have primarily been developed for legacy versions of Qiskit (0.x), limiting their applicability to current stable releases. This work presents QCRMut, a mutation testing tool for quantum circuits compatible with stable Qiskit versions, which supports controlled preservation of circuit structure through immutable positions and enables efficient, representative random mutant generation as an alternative to exhaustive mutation.

We develop QCRMut according to four design principles: unicity, similarity, representativity, and coverability. We evaluate the tool empirically by comparing mutation scores obtained from randomly sampled mutant subsets against those produced by exhaustive mutation over a benchmark suite of quantum circuits. We further analyse sensitivity to random seeds and apply statistical tests to assess the robustness of the observed differences. 

Finally, we compare QCRMut with existing mutation testing tools. Across the benchmark suite, randomly generated mutant subsets produced by QCRMut achieve mutation scores that are comparable to those obtained via exhaustive mutation. The results are stable across different random seeds and highlight practical limitations in existing approaches that are addressed by our method.

Overall, QCRMut provides a practical and extensible solution for mutation testing of quantum circuits by combining compatibility with stable Qiskit releases, controlled structure preservation, statistically sound evaluation, and efficient mutant generation. In addition, it enables the mutation and analysis of circuits that cannot be handled by previous tools.

\textbf{Keywords:} Formal Methods, Mutation Testing, Quantum Computing.
\end{abstract}

\textit{\footnotesize This work was supported by a predoctoral contract funded by the Universidad Complutense de Madrid and Banco Santander (CT25/24), the State Research Agency (AEI) of the Spanish Ministry of Science and Innovation under grant PID2021-122215NB-C31 (AwESOMe), and the Comunidad de Madrid under grant TEC-2024/COM-235 (DESAFíO-CM). This work has been submitted to the IEEE for possible publication. Copyright may be transferred without notice, after which this version may no longer be accessible.}

\vspace{-10pt}
\section{Introduction}
\label{S1:Intro}

Quantum computing is based on the principles of quantum mechanics to create a new paradigm in computing, distinct from classical computer engineering. Unlike classical computers, which use bits, quantum computers use quantum bits, or qubits. Qubits have different properties, such as superposition and entanglement, which lead to the desired quantum advantage. Quantum circuits are the framework within which quantum computations, also known as quantum gates, are performed. A quantum circuit consists of sequences of quantum gates that manipulate qubits to change the state.

Testing is a critical process in software development aimed at increasing confidence in a program functional correctness and its compliance with specific requirements. Dynamic testing involves executing software with the intent of finding errors and verifying that it behaves as expected under various conditions. A widely used testing technique in classical computing is \emph{mutation testing}.

Mutation testing is a technique where small changes are introduced into the code to represent potential errors. A test suite is then executed to determine whether it can detect these mutants. One of the problems we face in this field is the existence of equivalent mutants. These are syntactically different programs that yield the same result as the program under test. Significant work has been done in this field over the years, and this has been collected by Papadakis et al. in their 2019 study~\cite{papadakis2019mutation}. Later developments in the area include work combining mutation testing with LLMs~\cite{hassan2024llm}, as well as applications in smart contracts~\cite{li2019musc,honig2019practical} and deep learning~\cite{ma2018deepmutation,li2025drlmutation}. Sánchez et al. reviewed how mutation testing is applied in practice in their 2024 work~\cite{sanchez2024mutation}.

However, mutation testing is computationally intensive due to the large search spaces generated by mutant operators and the variety of inputs. Some attention has been given to reduction techniques, summarised in the 2010 work of Usaola et al.~\cite{usaola2010mutation} and in a systematic review by Pizzoleto et al.~\cite{pizzoleto2019systematic}. They group these techniques into three general approaches: \emph{do faster}, \emph{do fewer} and \emph{do smarter}. Our aim is to collect these techniques and reduce the cost of mutation testing, specifically in the context of quantum computing, where it has received limited attention.

The remainder of the paper is structured as follows. In Section~\ref{S2:Related}, we review the current state of the art and outline the starting point of our research. Section~\ref{S3:ToolAbs} introduces our RQs and the theoretical development of QCRMut (architecture and design decisions), followed by Section~\ref{S4:ToolDev}, where we adapt these decisions to implement the tool. Furthermore, Section~\ref{S5:Exp} presents our experiments, identifies suitable settings for random mutation, and discusses threats to validity. Finally, Section~\ref{S6:Conclusion} concludes the paper.

\vspace{-10pt}
\section{Related work}
\label{S2:Related}

In this section, we review the different testing techniques that have been introduced in quantum computing. Hoare's logic~\cite{ying2012floyd} and assertions were the first approaches used by researchers to develop testing techniques in quantum computing. This is illustrated by the work presented by Huang et al.~\cite{huang2019statistical}, where the authors propose the idea of using assertions as a debugging tool for quantum programs. Building on this idea, there was a subsequent evolution to runtime assertions. Zhou et al.~\cite{zhou2019quantum} propose the key idea of enabling dynamic assertions by introducing ancilla qubits. Later, the same authors~\cite{liu2020quantum} expanded their experiments. Finally, Li et al. proposed Proq~\cite{li2020projection}, a projection-based runtime assertion scheme in which operators generate projection-based predicates, providing flexibility in their placement.

All these previous works influenced Honarvar et al. to create a tool named QSharpCheck~\cite{honarvar2020property}. They present this approach as a property-based framework tailored for $Q\#$. Then, the study of input--output coverage criteria by Ali et al.~\cite{ali2021assessing,wang2021quito} constitutes a valuable continuation. The authors call it the Quito approach, where they define three coverage criteria according to their focus: input, output, and input--output. One of the key contributions of this paper is their formal definition of important elements such as quantum program, specification, and input or valid output values. We can observe combinatorial testing by Wang et al.~\cite{wang2021application}, where QuCAT tries to generate test suites with different classical generation algorithms, evaluating the effectiveness against random test suite generation.

Metwalli et al. introduced one of the first approaches to a debugging tool in~\cite{metwalli2022tool}, proposing the slicing of quantum programs to test each slice individually. Metamorphic testing has also been explored in this field, such as in the study of software systems by Paltenghi et al.~\cite{paltenghi2023morphq}, as well as in quantum programs by Abreu et al.~\cite{abreu2022metamorphic} and Costa et al.~\cite{costa2022asserting}.

Other approaches include the use of automata by Chen et al.~\cite{chen2023autoq} and static analysis of quantum programmes~\cite{zhao2023qchecker, paltenghi2024analyzing}. Further developments can be found in several surveys, such as those by Serrano et al.~\cite{serrano2022quantum}, García et al.~\cite{garcia2023quantum}, and the 2025 \emph{Road to 2030} survey by Leite et al.~\cite{leite2025testing}.

Mutation testing is a fault-based testing technique where small modifications (mutations) are introduced into a program to create mutant versions. These mutants are then tested to check if existing test cases can detect the faults. This technique helps in evaluating the effectiveness of test cases. When applied to quantum computing, mutation testing has been adapted to extend the classical approach to the properties of quantum programs. Some works already mentioned previously, such as~\cite{wang2021quito,honarvar2020property}, use mutation testing as a final resort to answer their hypotheses and research questions.

Wang et al.~\cite{wang2021qdiff} introduced quantum mutant operators to assess quantum platforms instead of quantum programs, obtaining some interesting results and bugs. Another work~\cite{wang2022mutation} is also based on mutation testing but focuses mainly on analysing the likelihood of having an equivalent mutant. The authors define a discount factor based on how many previous tests have not killed the mutant, which helps in determining the possibility of it being an equivalent mutant.

Some progress has been made towards establishing mutation testing techniques for quantum software. Two notable approaches discussed in the literature are Muskit by Mendiluze et al.~\cite{mendiluze2021muskit} and QMutPy by Fortunato et al.~\cite{fortunato2022mutation}, both of which have been explored in combination with Qiskit. Muskit performs source-level analysis and employs an exhaustive procedure to generate all possible mutations at each identified gap in the quantum circuit. The resulting mutants are then executed, and their outputs are evaluated using an oracle, following the methodology proposed in earlier studies~\cite{ali2021assessing,wang2021quito,wang2021qdiff}.

QMutPy, in contrast, extends the classical mutation testing framework MutPy by incorporating quantum-specific mutation operators and a notion of gate equivalence, implemented through Python’s Abstract Syntax Tree. Mutants are processed and analysed via code assertions, and the authors demonstrate that refining these assertions can lead to improved mutation scores. In 2025, Mendiluze et al.~\cite{mendiluze2025quantumBench} present a comprehensive, large-scale empirical study on quantum circuit mutation testing, involving more than 700,000 faulty benchmarks generated using Muskit from 382 real-world quantum circuits. Their results provide valuable evidence regarding the behaviour of quantum mutants and offer insights for the systematic development and evaluation of future quantum mutation testing techniques.

To date, these tools have been developed using preliminary versions of
Qiskit, employing an exhaustive approach to mutant generation. In this
work, we aim to advance the state of the art by introducing a random
generator tool built upon a stable version of Qiskit, designed to
reduce the number of mutants requiring execution and, consequently, to
lower computational costs. Furthermore, we propose novel concepts
relating to gate coverage and the understanding of a quantum circuit
as an entity, providing mechanisms to preserve substructures, such as oracle inputs and subroutines. We are also
integrating an initial automated testing framework designed to
facilitate the mutation testing process.

\section{Tool Abstraction}
\label{S3:ToolAbs}
This section outlines the concepts guiding our tool's development. Our primary aim is to utilise Qiskit's class structure to simplify the mutation process via high-level operations, eliminating the need to inspect the program's source code. This aligns our approach with other SDKs and QASM. We now introduce the concepts that shape the mutants structure.

\subsection{Insights and expectations}

We treat each quantum algorithm as a single entity, recognising that its particular structure may affect the mutation process. Some algorithms do not take the input at initialisation; instead, the input may appear as an oracle within the circuit. We therefore preserve the position of inputs and other key elements (e.g., subroutines), treating these positions as immutable. This is one of the aims in creating our tool, which is one of the main differences from previous approaches. A further goal is to generate correct-by-construction mutants and avoid stillborn mutants. However, the main difference between existing tools and ours is the mutation approach. While others generate an exhaustive set of mutants, we adopt a randomised approached and aim to produce a representative subset that yields similar results.

We build on the structure of Qiskit's \lstinline|QuantumCircuit| class and access circuit information via its attributes. In particular, \lstinline|data| exposes the circuit's ordered list of instructions, allowing us to modify the circuit structure and, if required, individual instructions. This process is detailed in the following subsections.

We previously developed an initial version of this tool for Qiskit 0.x, which is available in our repository. A key advance in this work is that our tool adapts to, and fully supports, stable Qiskit versions (Qiskit 1.3 and later). The transition from Qiskit 0.x to Qiskit 1.x introduced significant changes, including a Rust implementation of circuit instructions and updates to the methods for creating and adding gates to quantum circuits. These changes are explored in the following sections of this paper.

To align our research with these aims, we focus on the following research questions. \textbf{RQ1} is the primary objective; the others guide the analysis:

\vspace{-3pt}
\begin{description}
    \item[\textbf{RQ1}:] Do our randomly generated mutants represent those generated exhaustively?
    \item[\textbf{RQ2}:] Which of the proposed random approaches is the closest to the exhaustive method?
    \item[\textbf{RQ3}:] Is there a statistically significant difference between our random approach and the exhaustive mutation results?
    \item[\textbf{RQ4}:] How does the chosen seed affect the mutation score?
    \item[\textbf{RQ5}:] How does this tool compare with existing tools?
\end{description}

We use mutation score as the primary metric to answer these questions. Before presenting the experiments, we describe the tool and its structure.

\subsection{Definitions and basic intuition}

After defining some of our tool insights and identifying its unique features, we have highlighted its ability to preserve a specific position in a quantum circuit without undergoing mutation. To implement this feature, we considered two different approaches. One option was to add an attribute to the \lstinline|QuantumCircuit| class to indicate the immutable positions within the instruction list. However, this approach would require modifications whenever mutations occur.

Instead, we chose a different path. We initially aimed to create an abstract class, \lstinline|PlaceHolder|, as a child of \lstinline|CircuitInstruction|. This would allow us to introduce immutable elements into our circuits and assign custom names for easier instantiation. This approach preserves the integrity of \lstinline|QuantumCircuit| by treating the placeholder positions like any other instruction, while excluding them from the mutation set. However, in Qiskit 1.x, the \lstinline|CircuitInstruction| class has been marked as a final class in Python due to its new implementation. Consequently, we adapted our approach: instead of using an abstract class, we implemented a method under an analogous name. This method generates an empty circuit instruction at the desired position with the specified name, maintaining the full intended functionality. We note the importance of the name provided, as it acts as the key for swapping the placeholder with the desired quantum subroutine.

Another principle guiding our tool is to generate only mutated quantum
circuits that are correct by construction. Let us first provide a
couple of definitions that will help us achieve this goal. Then, we
will need to work with our mutation operators and ensure that once
they are applied, the produced circuit is correct.

First, we define the set of gates supported by our tool. We use 38
gates, collected in \lstinline|mutableGateSet|. These gates were
selected from a combination of the those used in
QMutPy~\cite{fortunato2022qmutpy} and those in the Qiskit equivalence
library. For example, \lstinline|gms| is a valid in QMutPy but is no
longer included in the Qiskit equivalence library; we therefore
exclude it.

Due to changes in Qiskit 1.x, and for simplicity,
we consider only gates that can be added to a quantum
circuit by calling the corresponding method by name,
\lstinline|qc.{name}()|. This choice relates to our mutant-generation
methodology.

Each gate has different parameters. We focus on two properties: the number of qubits to which the gate applies and the number of parameters it takes (e.g., rotation angles). We use the following notation: for a gate $g$, $\numQubits(g)$ and $\numParam(g)$ denote its qubit and parameter arity, and $\name(g)$ denotes its name. Examples are given in Table~\ref{tab:ParEx}.

\begin{definition}{\textbf{Mutable gate set}}: Let us denote QCRMut mutable gate set as:
    \begin{align*}
        \MutableGateSet \coloneqq \{&\text{x, h, z, y, t, sx, sdg, s, tdg, sxdg, id, p,}\\
        &\text{rz, ry, rx, r, u, swap, iswap, dcx, cz, }\\
        &\text{cy, cx, ch, cs, csx, csdg, rzz, rzx, ryy,}\\
        &\text{rxx, crz, cry, crx, cp, cswap, ccx, ccz}\}
    \end{align*}
\end{definition}

\begin{table}
    \centering
    \begin{tabular}{|c||c|c|c|}
        \toprule
         Gate & \name() & \numQubits() & \numParam() \\
         \midrule
         XGate & x & 1 & 0 \\
         \midrule
         CSGate & cs & 2 & 0 \\
         \midrule
         UGate & u & 1 & 3 \\
         \midrule
         RZZGate & rzz & 2 & 1 \\
         \bottomrule
    \end{tabular}
    \caption{Gate functions example.}
    \label{tab:ParEx}
\end{table}

Before we explain our mutation operators, we introduce gate equivalence, which we understand as \emph{syntactic} equivalence. The concept is similar to that introduced by Fortunato et al.~\cite{fortunato2022qmutpy}. We formalise this below.

\begin{definition}{\textbf{Gate equivalence}}\label{Def:ClassEqui}: Let $\text{p},\text{q}\in \MutableGateSet$, and $\sim$ the gate equivalence relation where:
    \begin{align*}
        \text{p}\sim \text{q} \iff \begin{split}
            \big(\numQubits(\text{p}) &=\; \numQubits(\text{q})\big) \\
            &\;\;\;\land \\
            \big(\numParam(\text{p}) &=\; \numParam(\text{q})\big)
        \end{split}
    \end{align*}
  \end{definition}

It is straightforward to see that $\sim$ is reflexive, symmetric, and transitive; therefore, it is an equivalence relation. This notion allows us to exchange gates without changing their qubit or parameter arity, avoiding changes to qubits or parameters and thereby maintaining circuit correctness.

\begin{proposition}
    Let $\sim$ be the relation defined in Definition~\ref{Def:ClassEqui}, and let $f:\;\MutableGateSet\;\to\mathbb{N}\times\mathbb{N}$, such that:
    \begin{equation*}
        f(\text{g})=(\;\numQubits(\text{g}),\numParam(\text{g})\;)
    \end{equation*}
    Then the quotient set of gate–equivalence classes is
    \begin{equation*}
        \MutableGateSet/\!\sim\; = \big\{\,f^{-1}((n,m)) : (n,m)\in f(\;\MutableGateSet\;)\,\big\}.
    \end{equation*}
    In other words, each equivalence class $[g]_\sim$ is exactly the preimage,
    \begin{equation*}
        [\text{g}]_\sim = f^{-1}\!\big(f(\text{g})\big)
    \end{equation*}
    Thus the classes denoted \texttt{nqmp} are precisely the nonempty sets $f^{-1}((n,m))$.
\end{proposition}

\begin{proof}
    From the definition of $\sim$ we have
    \begin{equation*}
        \text{p}\sim \text{q} \iff f(\text{p})=f(\text{q})
    \end{equation*}
    Hence the equivalence class of any g is
    \begin{equation*}
        [\text{g}]_\sim = \{\text{h} : \text{h}\sim \text{g}\} = \{\text{h} : f(\text{h})=f(\text{g})\} = f^{-1}(f(\text{g}))
    \end{equation*}
    Collecting all such classes yields the quotient $\MutableGateSet/\!\sim$, which is exactly the set of preimages $f^{-1}((n,m))$ for the pairs occurring in~$f$ image.
\end{proof}

These equivalence classes are summarised in Table~\ref{tab:gates}. Some differences arise compared with QMutPy~\cite{fortunato2022qmutpy}. For instance, QMutPy considers cswap equivalent to crz, despite the fact that they differ in both the number of qubits and the number of parameters.

\begin{table}[ht]
\begin{center}
\resizebox{\columnwidth}{!}{%
  \begin{tabular}{|c|c||c||l|}
    \toprule
    \#Qbits & \#Param & Class Name& Gates\\
    \midrule
    \midrule
    1& 0 & 1q0p & \{x, h, z, y, t, sx, sdg, s, tdg, sxdg, id\} \\
    1& 1 & 1q1p & \{p, rz, ry, rx\} \\
    1& 2 & 1q2p & \{r\}  \\
    1& 3 & 1q3p & \{u\} \\
    2& 0 & 2q0p & \{swap, iswap, dcx, cz, cy, cx, ch, cs, csx, csdg\} \\
    2& 1 & 2q1p & \{rzz, rzx, ryy, rxx, crz, cry, crx, cp\} \\
    3& 0 & 3q0p & \{cswap, ccx, ccz\} \\
  \bottomrule
\end{tabular}%
}
\end{center}
\caption{Gate equivalence classes.}
\label{tab:gates}
\end{table}

Before concluding this subsection, we introduce one final concept. Having identified the set of mutable gates, we summarise the pertinent information from the quantum circuit data to facilitate mutation. In particular, we record the positions of mutable elements within the circuit under test.

\begin{definition}{\textbf{Mutable index list}}\label{Def:MutIndList}: Let $QC$ be the quantum circuit under test, $Data^{QC}$ the ordered list of gates of $QC$ according to Qiskit, $N\in\mathbb{N}$ the length of $Data^{QC}$. We denote the mutable index list as $\mathcal{I}$, where:
    \begin{equation*}
        \mathcal{I} \coloneqq \{i \in \mathbb{N}, 0 \leq i < N\;|\; \name(Data^{QC}[i]) \in \MutableGateSet \}
    \end{equation*}
\end{definition}

Having established the structure of the \lstinline|placeHolder|, defined the set of mutable gates, and identified the equivalent classes, we now turn our attention to the process of creating mutants. We begin by examining the mutation operators, then discuss the incorporation of randomness in mutant generation, and finally present a new definition of gate coverage.

\subsection{Mutantion operators}

We use a reduced set of mutation operators. Previous work proposes additional operators \cite{mendiluze2021muskit,fortunato2022qmutpy}; however, we focus on mutations of the main circuit and exclude measurement mutations so that we remain in the deterministic part of the quantum program. Therefore, we ignore any operator that mutates measurement gates. The operator characteristics are as follows:

\begin{itemize}
    \item \textbf{Insertion operator}: This operator allows the insertion of a gate from our gate set at any position in the circuit. It is used as a ``joker'' operator for the extreme exceptions, like not having any mutable gates.
    \item \textbf{Deletion operator}: This operator removes the corresponding circuit instruction from the circuit, based on the list provided by \lstinline|data|.
    \item \textbf{Gate name operator}: This operator swaps the corresponding gate with a new equivalent but different gate, based on its equivalence class.
    \item \textbf{Gate attributes operator}:  This operator modifies the attributes, qubit application, and parameters of the gate, as appropriate.
\end{itemize}

All these operators represent common errors that programmers may encounter while coding. Let us see some visual examples of these operators in Table \ref{tab:OpExample}.

\begin{table}[H]
  \begin{center}
\newsavebox{\circins}
\newsavebox{\circdel}
\newsavebox{\circori}
\newsavebox{\circname}
\newsavebox{\circattr}
\savebox{\circori}{
  \begin{quantikz}
    & \gate{H} & & \\
    & \gate{H} & \gate{X} &
  \end{quantikz}}
\savebox{\circins}{
  \begin{quantikz}
    & \gate{H} & \gate{Z} &  \\
    & \gate{H} & \gate{X} &
  \end{quantikz}
}
\savebox{\circdel}{
  \begin{quantikz}
    & \gate{H} & \\
    & \gate{X} &
  \end{quantikz}
}
\savebox{\circname}{
  \begin{quantikz}
    & \gate{H} & \qw & \\
    & \gate{Z} & \gate{X} &
  \end{quantikz}
}
\savebox{\circattr}{
  \begin{quantikz}
    & \gate{H} & \gate{X} & \\
    & \gate{H} & \qw &
  \end{quantikz}
}
\resizebox{\columnwidth}{!}{%
  \begin{tabular}{|l|c|cccc|}
    \toprule
    Op & Original QC & Insert & Delete & Name & Attribute\\
    \midrule
    QC &
    \usebox\circori&
    \usebox\circins&
    \usebox\circdel&
    \usebox\circname&
    \usebox\circattr
    \\
    \bottomrule
  \end{tabular}%
}
\end{center}
\caption{Mutant operators example.}
\label{tab:OpExample}
\end{table}

Let $\Op = \{\Del,\:\Name,\:\Attr,\:\Ins\}$ denote the operator set, and let $\Op^- = \Op\setminus\{\Ins\}$ denote the reduced operator set.

There are certain cases in which we cannot mutate a position using a
particular operator. To capture this, we define an \(\Exc\) function that identifies such situations and will assist us in later definitions.

\begin{definition}{\textbf{Exception function}}: Let $QC$ be the quantum circuit under test, $Data^{QC}$ the ordered list of gates of $QC$ according to Qiskit, and $n\in\mathbb{N}$ the length of $Data^{QC}$. Then, $\Exc: \Op^- \times\;[0,n) \rightarrow\; \{True,\;False\}$, such that:
    \begin{equation*}
        \Exc(op,i) \coloneqq  \begin{dcases*}
          False & if $op = \Del$ \\
          \Exc_{Name}(i) & if $op = \Name$ \\
          \Exc_{Attr}(i) & if $op = \Attr$ \\
       \end{dcases*}
    \end{equation*}
    Where,
    \begin{align*}
        \Exc_{Name}(i) &\coloneqq \Bigl|\bigl[\name(Data^{QC}[i])\bigr]_{\sim}\Bigr| = 1\\
      \Exc_{Attr}(i) &\coloneqq \Big(\numQubits(QC) = 1 \;\;\land \\
                     &\qquad \;\numParam(Data^{QC}[i]) = 0\Big)\\
                     & \hspace{60pt}\;\lor\; \\
                     &\qquad \Big(\numQubits(QC) = 2\;\;\land \\
                     &\qquad \;\;\name(Data^{QC}[i]) = \text{swap}\Big)
    \end{align*}
\end{definition}

These exceptions follow directly from the operator definitions: we
cannot vary the name of a gate when no alternative name exists, and
the $\Attr$ operator cannot be applied when neither qubits nor
parameters may be modified.

This leads to a common issue in mutation testing: the generation of equivalent mutants, such as deleting an identity gate or swapping identical qubits in a swap gate. The equivalence problem will be handled as a tool-development concern; the definition above addresses only the inherent impossibility of mutation.

\subsection{Random generator approaches}

Up to this point, we have presented several definitions concerning our
set of gates, their properties, and the mutant operators. However, we
are aiming to apply randomness to the mutant-generation process, and
we have not yet discussed how this will be incorporated. Firstly, let
us introduce the principles we follow:

\begin{itemize}
    \item \textbf{Unicity}: The generation procedure should avoid producing the same mutant more than once.
    \item \textbf{Similarity}: The randomly generated subset of mutants should
      resemble, and be representative of, the full exhaustive set.
    \item \textbf{Representativity}: All operators should be appropriately
      represented within the selected subset of mutants.
    \item \textbf{Coverability}: All positions within the quantum circuit should be mutated.
\end{itemize}

A straightforward approach would be to generate mutants at random
while checking for duplicates. However, this forces us to decide how
we choose an operator in the first place and what probabilities we
assign to each, while still preserving similarity to the exhaustive
procedure.

To clarify this issue, consider the operators and the number of
mutants each can produce. The insertion operator depends solely on the
position within the circuit: it may insert at any location and may
introduce any gate from our gate set. In contrast, the remaining
operators, $\Op^-$, depend both on the ability to mutate a particular
position and on the properties of the gate currently occupying that
position. For instance, the deletion operator can produce at most as
many mutants as there are gates in the circuit.

This implies that the insertion operator will always produce the
highest number of mutants. Therefore, to maintain representativity, we
will not apply probabilistic selection ad-hoc over the full mutant data set. This
choice will be fully motivated in Section~\ref{Sec4.2:BuildingB}, once we
present our quantum circuits under test and the mutants produced
exhaustively.

Therefore, our strategy is to generate all potential mutants for every
operator $op \in \Op^-$, and thereafter select a random mutant subset
from these complete sets.
For the insertion operator, however, we must
decide whether to adopt the same strategy—e.g., generate all insertion
mutants in advance—or instead generate them one at a time while
checking for duplicates. To ensure efficiency, QCRMut first identifies the space of all
possible mutations, operator plus parameters, and then samples from this
space to generate the actual mutant circuits, avoiding the overhead of
creating unused circuits. 

This constitutes the first decision in our methodology, which we
denote by $\allins$: whether insertion mutants should be
pre-generated in full or generated incrementally.

Earlier, we emphasised the importance of avoiding duplicate
mutants. Although duplicate detection can be performed during
generation, or avoided entirely by enumerating all mutants in advance,
this leads to a broader question: what constitutes the “same” mutant?
In this discussion, we refer strictly to syntactic equivalence, as defined in mutation testing theory.

Since the quantum circuit is represented as a data list, we consider
two mutants structurally equivalent if they apply the same operator,
with the same parameters, to the same circuit position. Such
duplicates are automatically avoided by adhering to the generation
procedures described above.

Even so, different mutants may still yield identical graphical
representations and behaviour, according to our earlier
definitions. This arises from the positioning of gates within the
quantum circuit. Let us illustrate this with Example~\ref{Ex:simplifyData}.

\begin{example}
    \label{Ex:simplifyData}
    Let us observe the original quantum circuit, QC, their simplified data, and the mutated circuits in Table~\ref{tab:RedExample}. We can observe how inserting the same mutation but in different indexes, produce the same graphical quantum circuit, yielding the same result.
\end{example}

\begin{table}[H]
  \begin{center}
  \newsavebox{\cirQC} \savebox{\cirQC}{
        \begin{quantikz}
          \lstick{0} & \gate{X} & \ctrl{1} &    \qw   &    \qw   &    \qw   & \gate{X} & \\
          \lstick{1} & & \gate{X} & \gate{X} & \gate{H} & \gate{Z} &
          \ctrl{-1} &
        \end{quantikz}}

\newsavebox{\cirQCA}
\savebox{\cirQCA}{
  \begin{quantikz}
    \lstick{0} & \gate{X} & \ctrl{1} &    \gate{X}   &    \qw   &    \qw   & \gate{X} & \\
    \lstick{1} &          & \gate{X} & \gate{X} & \gate{H} & \gate{Z} & \ctrl{-1} &
  \end{quantikz}
}
\resizebox{\columnwidth}{!}{%
  \begin{tabular}{|c|c|}
          \toprule
          $Data^{QC}$ & $ \Big[\big(\text{x},(0)\big),\big(\text{cx},(0,1)\big),\big(\text{x},(1)\big),\big(\text{x},(1)\big),\big(\text{x},(1)\big),\big(\text{cx},(1,0)\big)\Big]$ \\
          \midrule
          QC &
          \usebox\cirQC \\
          \midrule
          \midrule
    $Data^{m_1}$ & $ \Big[\big(\text{x},(0)\big),\big(\text{cx},(0,1)\big),\big(\text{\textbf{x}},(\text{\textbf{0}})\big),\big(\text{x},(1)\big),\big(\text{h},(1)\big),\big(\text{z},(1)\big),\big(\text{cx},(1,0)\big)\Big]$ \\
    \midrule
    $Data^{m_2}$ & $ \Big[\big(\text{x},(0)\big),\big(\text{cx},(0,1)\big),\big(\text{x},(1)\big),\big(\text{\textbf{x}},(\text{\textbf{0}})\big),\big(\text{h},(1)\big),\big(\text{z},(1)\big),\big(\text{cx},(1,0)\big)\Big]$ \\
    \midrule
    $Data^{m_3}$ & $ \Big[\big(\text{x},(0)\big),\big(\text{cx},(0,1)\big),\big(\text{x},(1)\big),\big(\text{h},(1)\big),\big(\text{\textbf{x}},(\text{\textbf{0}})\big),\big(\text{z},(1)\big),\big(\text{cx},(1,0)\big)\Big]$ \\
    \midrule
    $Data^{m_4}$ & $ \Big[\big(\text{x},(0)\big),\big(\text{cx},(0,1)\big),\big(\text{x},(1)\big),\big(\text{h},(1)\big),\big(\text{z},(1)\big),\big(\text{\textbf{x}},(\text{\textbf{0}})\big),\big(\text{cx},(1,0)\big)\Big]$ \\
    \midrule
    $m_i$ &
    \usebox\cirQCA \\
    \bottomrule
  \end{tabular}%
}
\end{center}
\caption{Mutation examples without reduction.}
\label{tab:RedExample}
\end{table}

This leads to the second decision in our approach, whether we base
equivalence solely on the circuit’s syntactic data, or whether we also
apply a reduction step that considers graphical equivalence; we denote
this as $\reduction$. As both approaches have their own rationale, one
may argue that such mutants represent several distinct faults and
should therefore retain that weight, or alternatively that they should
be treated as a single mutant because they behave identically.

The resulting option space is given by the following combinations:

\begin{itemize}
\item Reduction with all insertion data pre-generated.
\item Reduction without pre-generating all insertion data.
\item No reduction with all insertion data pre-generated.
\item Neither reduction nor full pre-generation of insertion data.
\end{itemize}

We have now observed most of our principles in a preliminary form. However, let us look more deeply into the coverability principle.

\subsection{Coverage and analysis}

When it comes to testing, one of the key concepts we rely on is coverage analysis of the code. In classical computer engineering, it is common to measure the lines of code (LOC) covered during the mutation process. Previous works such as Muskit \cite{mendiluze2021muskit} and QMutPy \cite{fortunato2022qmutpy} have adopted this principle in their approaches. Their methods are exhaustive, aiming to cover either all gaps or all nodes of the Python syntax tree, utilising the structure of MutPy. However, in our case, as we are mutating the \lstinline|QuantumCircuit| class, we will assess the coverage based on the number of gates rather than the LOC, which is common in quantum computing.

We would like to study how many mutants we need to generate to achieve
full coverage. Let us consider full gate coverage when each gate is
mutated at least once by each possible operator, and each gap
is mutated at least once by the insertion operator.
We identify a gap as an element of the set ${\cal I}$ (Definition~\ref{Def:MutIndList}).

\begin{definition}{\textbf{Full gate coverage}:}\label{Def:FullCoverage} Let QC be the quantum circuit under test, $\mathcal{I}$ the QC mutable index list, $\mathcal{M}$ a set of mutants of QC,  and $\:\forall i \in\mathbb{N}\; \;0 \leq i \leq N$, $\forall\:op\in \Op\setminus\{\Del\}$,
    \begin{align*}
        \mathcal{M}_{op}^i  &\coloneqq \{m \in \mathcal{M}\;| \;Data^m[i]\text{ has been mutated}\\
                            &   \hspace*{7em}\text{with operator }op\}. \\
        \mathcal{M}_{Del}^i  &\coloneqq \{m \in \mathcal{M}\;| \;Data^{QC}[i]\text{ deleted in }m\}.
    \end{align*}
    Then, $\mathcal{M}$ \emph{has full coverage} iff
    \begin{align*}
        \begin{split}
        \big(\;\forall i \in \mathcal{I} \;\;\;\forall op \in Op^*&,\;\Exc(op,i)\;\vee\;\mathcal{M}_{op}^i \neq \varnothing\big) \\
        &\land \\
        \;\;\big(\forall i \in \mathbb{N}, \;0 \leq&\:i \leq N, \;\;\mathcal{M}_{Ins}^i \neq \varnothing\big)
        \end{split}
    \end{align*}
\end{definition}

To estimate the number of mutants, we relate this problem to the \emph{Coupon Collector's} problem~\cite{flajolet1992birthday}. For simplicity, we bound the number of mutable instructions by the number of gates in the circuit, denoted $N$. We aim to collect $N+1$ \emph{coupons} from the insertion operator (e.g., all insertion gaps) and, at most, $N$ \emph{coupons} from the remaining operators. Each coupon group has a different probability. However, the insertion group has probability of at least $1/(N+1)$. Our goal is to obtain all coupons from each group, which corresponds to mutating each position. For simplicity, we will focus on a single question: How many mutants do we need to generate using the insertion operator to obtain the $N+1$ \emph{coupons}?

The solution to this question aligns with the equiprobable Coupon Collector's problem for $n$ coupons.  It is important to remember that this solution represents the expected number of executions needed. It is the mean of the distribution representing all possible values for a full collection:

\begin{equation}
    \exeins(n)= n\cdot H_n = n\cdot  \sum_{i=1}^n \dfrac{1}{i}
\end{equation}

Using Maclaurin–Cauchy test, we can bound $H_n$:
\begin{equation}
    \ln(n)+\frac{1}{n} \leq H_n \leq \ln(n) +1
\end{equation}

If we consider the upper bound and $n= N+1$ we obtain:
\begin{equation}
  \exeins = (N+1) \cdot \lceil \ln(N+1) +1\rceil
\end{equation}

Therefore, to achieve the desired coverage, we need to use the insertion operator $(N+1) \lceil \ln(N+1) +1\rceil$ times, where $N$ represents the number of instructions in the circuit. However, we decided that full coverage had to include obtaining all desired coupons, including the ones for modifying or deleting existing gates. As the insertion operator dominates the remaining operators, we will use this as an upper bound. Then, we will support the following number of executions.

\vspace{-10pt}
\begin{equation}\label{eq:minMutants}
    \exe = 4\:(N+1) \lceil \ln(N+1) +1\rceil
\end{equation}

We could conclude, then, that with $4(N+1) \lceil \ln(N+1) +1\rceil$
executions, we fulfil the desired coverage, as this serves as an upper
bound for the solution of the problem involving other operators, which
is $N\cdot H_N$. However, we should recall that the solution to the
Coupon Collector's problem provides the expected value and does not
ensure it. Despite that, we are going to consider such value as a good
approximation. As we are considering this as a good approximation, we
will use experiments to strengthen this hypothesis.

\section{Tool development}
\label{S4:ToolDev}

The main principles and definitions for our tool have already been
introduce. We now describe their implementation and show how the tool adheres to these principles.

This section builds from core components to the complete tool, relating each step to the definitions introduced earlier. It is organised into three parts: circuit preparation, circuit mutation, and circuit execution.

We use the Bernstein--Vazirani (BV) algorithm as a running example throughout this section. Table~\ref{tab:BVAlgorithm} shows the six-qubit circuit. Here, the algorithm solves an oracle problem and the oracle occupies a fixed position in the circuit; we include it explicitly in both the source code and the diagram.

\begin{table*}
\begin{center}
    \newsavebox{\cirBVI}
    \newsavebox{\cirlstBVI}
    \savebox{\cirBVI}{
        \begin{quantikz}
            \lstick{$q_0$} &\gate{H} & \qw & \gate[7,disable auto
        height]{\verticaltext{ORACLE}}  & \gate{H} &\\
            \lstick{$q_1$} &\gate{H} & \qw &  &\gate{H} &\\
            \lstick{$q_2$} &\gate{H} & \qw &  &\gate{H} &\\
            \lstick{$q_3$} &\gate{H} & \qw &  &\gate{H} &\\
            \lstick{$q_4$} &\gate{H} & \qw &  &\gate{H} &\\
            \lstick{$q_5$} &\gate{H} & \qw &  &\gate{H} &\\
            \lstick{$anc$} &\gate{H} &  \gate{Z} & &\gate{H} &
        \end{quantikz}}
    \begin{tabular}{p{15pt} @{} p{0.45\linewidth} |c @{}}
        \toprule
        & BV Alg source code & BV Alg quantum circuit \\
        \midrule
        &\begin{minipage}{\linewidth}
            \begin{lstlisting}[linewidth=\linewidth, frame = none, numbers=left, basicstyle=\small]
def BV_Alg(n: int = 6):
    qs = QuantumRegister(n, 'q')
    anc = QuantumRegister(1, 'anc')
    qc = QuantumCircuit(qs, anc)

    for i in range(n+1):
        qc.h(i)
    qc.z(anc)

    qc.append(ORACLE)

    for i in range(n):
        qc.h(i)

    return qc\end{lstlisting}
        \end{minipage} &
        \usebox\cirBVI \\
        \bottomrule
    \end{tabular}
\end{center}
\caption{BV Algorithm: 6 qubits.}
\label{tab:BVAlgorithm}
\end{table*}

\subsection{Quantum circuit preparation}

Before applying mutation, we must prepare the quantum circuit. One of the novel features provided by our tool is the ability to preserve specific positions in a quantum circuit so that they do not undergo mutation. We propose a method that allows us to insert an immutable circuit instruction. Listing~\ref{lst:placeholderDef} shows the function signature and the parameters that can be set for a typical instruction.

\vspace{5pt}
\begin{lstlisting}[caption=PlaceHolder method head., label={lst:placeholderDef}]
def placeHolder(qubits: list[Qubit], name: str = "Input",
   clbits: Optional[list[Clbit]] = None,
   param: Optional[list[AnyNum]] = None,
   label: Optional[str] = None) \
   -> CircuitInstruction
\end{lstlisting}

To ensure immutability, values from $MutableGateSet$ are not permitted for the \lstinline|name| parameter. Let us present an example of how to initialise the BV Algorithm.

\begin{example}
  {\textbf{BV Alg Set-Up}}: We want to preserve our input
  position so that we can later identify where to instantiate it. We
  will modify line 10 from Table~\ref{tab:BVAlgorithm} source
  code. The input applies to all qubits, which is the only necessary
  data in this case.

\begin{center}
    \lstinline|ORACLE = placeHolder(qc.qubits)|
\end{center}
\end{example}

It is important to provide the \lstinline|placeHolder| instruction with all necessary information, particularly the qubits to which it applies. Since these qubits will not be modified, they can affect the mutation process, especially in relation to our reduction concept. If incorrect qubits are specified, exceptions may arise when we attempt to instantiate the placeholder.

\subsection{Mutation phase I, building blocks}
\label{Sec4.2:BuildingB}

We begin with the building blocks of the tool. First, we explain how the concepts from the previous section are implemented. Next, we describe how each operator mutates the quantum circuit, including relevant Qiskit-specific considerations. We then discuss how many mutants should be generated to ensure coverability, and finally we review the reduction process.

The main concepts introduced in the previous section are implemented directly: $MutableGateSet$ as a set and the equivalence classes as a dictionary, where $nqmp$ is used as the key (with $n$ the number of qubits and $m$ the number of parameters). It is important to clarify the role of $MutableGateSet$: the tool uses this collection to identify which gates may be mutated, but it does not constrain which gates may be used when creating new mutants. Modifying this set may therefore introduce correctness issues or raise errors when the tool attempts to access missing information. New mutants are created using the equivalence dictionary.

We would like to compare our work with previous tools; however, differences in the underlying gate sets make direct comparison non-trivial. For this reason, our tool also supports exhaustive generation. This allows us to compare random and exhaustive results under the same conditions.

This introduces a new challenge: defining what \emph{exhaustive} means in the presence of gate parameters, which range over an infinite set. We therefore define a finite parameter set used by the tool, via a global variable $Exhaustive\_Parameters$. Our random mutation process is also based on this finite set, ensuring a well-defined mutant space.

\begin{definition}{$\mathbf{Exhaustive\_Parameters}$.}
    \begin{equation*}
    Exhaustive\_Parameters = \Bigg\{\dfrac{k\pi}{7},\;k \in\{1,2,...,6\}\Bigg\}
    \end{equation*}
\end{definition}

We have chosen this finite set of real numbers, as they do not produce equivalent rotations to other non-parametric gates. We have selected the number 7, because it is a prime number reasonably small. Avoiding an exponential growth in the number of parameters and $\pi\;/\;{2^n},\;n\in\mathbb{N}$ rotations. We did start in $k=1$, as we do not want to generate the identity with a $0$ degree rotation. However, this set can always be modified; doing so may result in equivalent mutants to the original circuit or equivalent rotations to other gates.

Let us now explain the challenges and how we approach the mutation of our quantum circuits. First, we focus on how we generate the required data, and then we describe how this mutation is incorporated into the circuit.

The data generation step could be the simplest part of the tool. Nevertheless, we took \emph{unicity} as one of our guiding principles. This principle shapes the data we generate. The following considerations reflect this principle, keeping in mind that in Qiskit the order of qubits is significant:

\begin{itemize}
    \item \textbf{Identity gate}: We do not mutate the identity gate, as it does not affect the quantum circuit. The only exception would be a name change; such a mutant is already generated by the insertion operator. Therefore, we remove the id gate from the $MutableGateSet$.
    \item \textbf{Swap gates}: The order of the qubits in a swap gate does not matter; hence, we generate only one of the possible permutations.
    \item \textbf{Double-controlled gates}: The order of the control qubits does not matter; therefore, only one configuration is generated.
    \item \textbf{Parameter value}: This was explained previously when the set of parameter values was introduced.
\end{itemize}

It is not possible to construct a multi-controlled gate with a swap gate as the target, since we consider only three-qubit gates, whereas such a construction would require at least four qubits. To ensure close adherence to our unicity principle, the complexity of qubit combinations is managed through a function dictionary in which the key represents the number of qubits involved, as shown in Listing~\ref{lst:FunDict}. This approach allows the combinations to be integrated using list comprehension during data generation, as illustrated in Listing~\ref{lst:dataInsEx}.

We now introduce the data generation procedure. Table~\ref{tab:MutDataOp} summarises the data that must be obtained for each operator.

\vspace{10pt}
\begin{lstfloat}
\begin{lstlisting}[caption=Function dictionary., label={lst:FunDict}]
funct_dict = {
    0: lambda name: [lambda q: q],
    
    1: lambda name: [lambda q: q]
      if name == 'swap'
      else  [lambda q: q,
          lambda q: (q[1],q[0])],
          
    2: lambda name: [lambda q: q,
          lambda q: (q[1], q[0], q[2]),
          lambda q: (q[2], q[0], q[1])]
      if name == 'cswap'
      else [lambda q: q,
          lambda q: (q[0], q[2], q[1]),
          lambda q: (q[1], q[2], q[0])]}
\end{lstlisting}
\end{lstfloat}

\begin{table}
    \centering
    \begin{tabular}{|c|c|ccc|}
        \toprule
         Operator & Index & Name & Parameters & Qubits \\
         \midrule
         Insertion & \cmark & \cmark & \cmark & \cmark \\
         Deletion & \cmark & \xmark & \xmark & \xmark \\
         Name & \cmark & \cmark & \xmark & \xmark \\
         Attributes & \cmark & \xmark & \cmark & \cmark \\
         \bottomrule
    \end{tabular}
    \caption{Mutation data per operator.}
    \label{tab:MutDataOp}
\end{table}

Let us analyse how the required data are obtained for each operator. This follows a specific selection order, since earlier choices may constrain later ones. First, all operators require an application index. In general, this index is selected from the mutable index list defined in Definition~\ref{Def:MutIndList}, except for the insertion operator, for which the index is chosen from the set of possible insertion positions (depending on the selected reduction strategy).The remaining data are selected as follows, depending on the mutation operator used:

\begin{itemize}
        \item \textbf{Name}: The Insertion operator obtains one name from the available possibilities; when generated at random, weighted selection is used. The Name operator selects a name from the equivalence class of the current name at the chosen index.
        \item \textbf{Parameters}: These are selected at random from $Exhaustive\_Parameters$. The number of parameters is determined by the name selected previously. They are represented as a tuple for comparison purposes.
        \item \textbf{Qubits}: These are selected at random from the $QC$. The number of qubits is determined by the name selected previously and is represented as an ordered tuple, for comparison purposes and, in this case, in accordance with Qiskit’s internal instruction representation.
\end{itemize}

All possible combinations of this information are used to generate the complete data set for each operator. Listing~\ref{lst:dataInsEx} illustrates an example of the insertion data generator. The final data for each operator are represented as a list of tuples containing the information specified in Table~\ref{tab:MutDataOp}, including an integer in position 0 that indicates the operator in use.

\vspace{5pt}
\begin{lstlisting}[caption= Insertion data generator fragment., label={lst:dataInsEx}]
insertParam = [(0, index, name, param, f(qubit_comb))
               ...
               for qubit_comb in Qubit_Combinations[num_qubits]
               ...
               for f in funct_dict[num_qubits](name)]
\end{lstlisting}

Once we have the mutant data, we must apply the mutation to the quantum circuit under test. This reveals a key difference between Qiskit versions (and therefore between tool versions). In Qiskit 0.x, it was sufficient to modify the relevant entry in the data. In stable versions of Qiskit, circuit instructions are immutable; therefore, we use \lstinline|to_mutable| and \lstinline|replace| to create a mutable copy of an operation and replace it at a given circuit position.

We briefly elaborate on a Qiskit-specific behaviour that required two implementation approaches. When we create a mutable copy of an instruction, we can update its parameters in place, but not its name. To change a gate name, we must create a new instruction. We include this detail as it helps explain the implementation choices and informs the reader.

Listing~\ref{lst:AttrOperator} shows the attribute-operator method. Here, we can modify parameters directly on the mutable instruction. By analogy, one might expect the name to be mutable too, adding \lstinline{aux_instruction.name = new_name}.

We illustrate this behaviour with a small example in Table~\ref{tab:NameExample}, using the same original quantum circuit. The mutations are as follows: both are applied at index 2; $m_1$ uses the Attribute operator with the qubit reassigned to 0, and $m_2$ applies the Name operator with the name Z.

\vspace{10pt}
\begin{lstfloat}
\begin{lstlisting}[caption= Gate attributes change
operator.,label={lst:AttrOperator}]
def attributesChange_Operator (
       testQC: QuantumCircuit,
       index: int,
       params: tuple[AnyNum,...],
       qubits: tuple[Qubit,...]) \
       -> QuantumCircuit:

    mutant = testQC.copy()
    
    # Create a mutable copy of
    # the instruction.
    aux_instruction = mutant \
      .data[index] \
      .operation.to_mutable()
      
    # Assign new parameters.
    aux_instruction.params = new_params
    
    # We replace the original
    # instruction with the new
    # parameters and qubit assignment.
    mutant.data[index] = \
      mutant.data[index] \
      .replace(operation = \
                  aux_instruction,
               qubits = new_qubits)
\end{lstlisting}
\end{lstfloat}
Although the data indicate that the mutation has been applied, the circuit representation does not change. Moreover, when we execute the mutant, it behaves as suggested by the diagram rather than by the data. We attribute this to Qiskit's newer implementation, in which an instruction is linked to the gate implementation and is not determined solely by its name after creation. We therefore modify the Name operator to create a new instruction with the required information (Listing~\ref{lst:NameOperator2}).

The remaining operators do not create unexpected issues, since they either delete an instruction or insert a new one. To conclude the building blocks, we analyse mutant calculation and how we enforce coverability.

Our first design decision is to enforce a minimum number of mutants, to avoid cases where a single mutant has an outsized impact on the mutation score. For example, with 100 mutants, each contributes 1\% to the score. We therefore require that no single mutant contributes more than 0.5\%, which we treat as a lower bound.

\begin{table}
  \begin{center}

  \newsavebox{\cirQCOr} \savebox{\cirQCOr}{
        \begin{quantikz}
          \lstick{0} & \gate{X} & \ctrl{1} &    \qw   &    \qw   &    \qw   & \gate{X} & \\
          \lstick{1} & & \gate{X} & \gate{X} & \gate{H} & \gate{Z} &
          \ctrl{-1} &
        \end{quantikz}
      }
\newsavebox{\cirQCB}
\savebox{\cirQCB}{
  \begin{quantikz}
    \lstick{0} & \gate{X} & \ctrl{1} &    \qw   &    \qw   &    \qw   & \gate{X} & \\
    \lstick{1} &          & \gate{X} & \gate{X} & \gate{H} & \gate{Z} & \ctrl{-1} &
  \end{quantikz}
}
\newsavebox{\cirQCC}
\savebox{\cirQCC}{
  \begin{quantikz}
    \lstick{0} & \gate{X} & \ctrl{1} &    \gate{X}   &    \qw   &    \qw   & \gate{X} & \\
    \lstick{1} &          & \gate{X} & \qw & \gate{H} & \gate{Z} & \ctrl{-1} &
  \end{quantikz}
}
\resizebox{\columnwidth}{!}{%
  \begin{tabular}{|c|c|}
    \toprule
    $Data^{QC}$ & $ \Big[\big(\text{x},(0)\big),\big(\text{cx},(0,1)\big),\big(\text{x},(1)\big),\big(\text{x},(1)\big),\big(\text{x},(1)\big),\big(\text{cx},(1,0)\big)\Big]$ \\
      \midrule
      QC &
      \usebox\cirQCOr \\
      \midrule
      \midrule
    $Data^{m_1}$ & $ \Big[\big(\text{x},(0)\big),\big(\text{cx},(0,1)\big),\big(\text{x},(\text{\textbf{0}})\big),\big(\text{h},(1)\big),\big(\text{z},(1)\big),\big(\text{cx},(1,0)\big)\Big]$ \\
    \midrule
    $m_1$ &
    \usebox\cirQCC \\
    \midrule
    \midrule
    $Data^{m_2}$ & $ \Big[\big(\text{x},(0)\big),\big(\text{cx},(0,1)\big),\big(\text{\textbf{z}},(1)\big),\big(\text{h},(1)\big),\big(\text{z},(1)\big),\big(\text{cx},(1,0)\big)\Big]$ \\
    \midrule
    $m_2$ &
    \usebox\cirQCB \\
    \bottomrule
  \end{tabular}%
}
\end{center}
\caption{Mutation operator examples, unexpected behaviour.}
\label{tab:NameExample}
\end{table}

To determine an appropriate minimum, we calculated the maximum number of mutants for the simplest circuit: a single qubit with no gates. This yields 286 mutants from the insertion operator alone. Since any circuit will generate at least this many insertion mutants, the minimum threshold must lie between 200 and 286; we therefore set it to 256.

\begin{landscape}
\begin{figure}
    \centering
    \scalebox{0.95}{
    \begin{tikzpicture}[node distance=2cm]

        \node (start) [startstop] {Receive Quantum Circuit};
        \node (iniDataGen) [process, below of=start, yshift=-1cm, text centered, text width=5cm] {Generate \Name, \Attr{} and \Del{} data};
        \node (exhaustive) [decision, right of=iniDataGen, xshift=3.5cm] {Exhaustive?};
        \node (AllInsExh) [process, below of=exhaustive, text centered, text width=5cm, yshift=-1cm] {Generate all \Ins{} data with Reduction};
        \node (MutNumber) [process, above of=exhaustive, yshift=1cm] {Compute number of mutants};
        \node (GenMut) [process, below of=AllInsExh] {Generate all mutants};
        \node (End) [startstop, left of=GenMut, text centered, text width=5cm, xshift=-3.6cm] {Save if required; Return mutants and data};
        \node(allIns) [decision, right of=MutNumber, xshift=3.7cm] {All\_Ins data?};
        \node(AllIns) [process, below of=allIns, text centered, text width=5cm, yshift=-4cm] {Generate all \Ins{} data};
        \node(SubSel) [process, below of=AllIns] {Mutant subset selection};
        \node(RandomIns) [process, right of=allIns, text centered, text width=5cm, xshift=3.8cm] {Generate \Ins{} data at random};
        \node(Enough) [decision, below of=RandomIns, yshift=-1cm] {Enough data?};

        \draw [arrow] (start) -- (iniDataGen);
        \draw [arrow] (iniDataGen) -- (exhaustive);
        \draw [arrow] (exhaustive.south) -- node[anchor=east] {Yes}  (AllInsExh);
        \draw [arrow] (exhaustive.north) -- node[anchor=east] {No}  (MutNumber);
        \draw [arrow] (AllInsExh) -- (GenMut);
        \draw [arrow] (GenMut) -- (End);
        \draw [arrow] (MutNumber) -- (allIns);
        \draw [arrow] (allIns) -- node[anchor=east] {Yes} (AllIns);
        \draw [arrow] (allIns) -- node[anchor=south] {No} (RandomIns);
        \draw [arrow] (AllIns) -- (SubSel);
        \draw [arrow] (SubSel) -- (GenMut);
        \draw [arrow] (RandomIns) -- (Enough);
        \draw [arrow] (Enough.south) -- ++(0,-1.63) -- node[anchor=south] {No} (AllIns.east);
        \draw [arrow] (Enough.east) -- ++(0.5,0) -- ++(0,-5) -- node[anchor=south] {Yes} (SubSel.east);
    \end{tikzpicture}}
    \caption{QCRMut workflow}
    \label{fig:WorkflowQCRMut}
\end{figure}
\end{landscape}

Table~\ref{tab:BVMutExample} reports the number of possible mutations per operator for our example, using a reduced exhaustive procedure. This behaviour extends to other circuits and, in practice, results in the insertion operator producing most mutants. We first compute the total number of mutants required to achieve the desired level of coverability. This calculation is driven by the insertion operator and used as the primary reference. The remaining question is how to distribute mutants across operators while supporting three principles: \emph{similarity}, \emph{representativity}, and \emph{coverability}.We analyse below how the chosen principle affects the allocation of mutants across operators:

\begin{itemize}
    \item \textbf{Similarity}: If the selected subset is intended to approximate the exhaustive mutant set, the insertion operator should account for nearly 100\% of the mutants.
    \item \textbf{Representativity}: If all operators are to be represented, and given that we aim to utilise all of them, a minimum number of mutants must be allocated to each operator.
    \item \textbf{Coverability}: To achieve full gate coverability, Definition~\ref{Def:FullCoverage}, each operator must be assigned at least the number of mutants required for this purpose.
\end{itemize}

\begin{lstfloat}
\begin{lstlisting}[caption= Gate name change operator.,
label={lst:NameOperator2}]
# We instantiate a new instruction,
# providing the new name.
aux_instruction = Instruction(new_name,
  mutant.data[index].operation.num_qubits,
  mutant.data[index].operation.num_clbits,
  mutant.data[index].operation.params)
# We replace the old instruction.
mutant.data[index] = mutant.data[index]\
  .replace(operation=aux_instruction)
\end{lstlisting}
\end{lstfloat}

\vspace{-10pt}
Each principle imposes different constraints; therefore, our objective is to provide a solution that satisfies these principles as closely as possible. We address this by defining a minimum percentage of mutants for each operator, which is then reviewed to ensure coverability. This approach preserves a large proportion of mutants for the insertion operator, thus maintaining similarity with the exhaustive set, while guaranteeing representativity across all operators and enabling coverability.

This minimum percentage is exposed as a user-configurable parameter. In our implementation, we set it to 5\%, which leaves approximately 85\% of the mutants assigned to the insertion operator.

The allocation follows two rules. If the assigned percentage exceeds the number of available mutants for a given operator, the remainder is reassigned to the insertion operator. Conversely, if additional mutants are required to meet coverability, we increase the total number of mutants so that the minimum percentage still satisfies the coverability requirement, preserving the similarity principle. Table~\ref{tab:BVMutExample} illustrates these calculations for BV.

\begin{table}[H]
        \centering
    \resizebox{\columnwidth}{!}{%
    \begin{threeparttable}
        \begin{tabular}{|c|ccccc|}
                \toprule
                Algorithm & \#Gates & \#Mut Gates & \multicolumn{1}{|c}{\#Mutants}& Initial & Final \\
                        \midrule
                BV & 15 & 14 & \multicolumn{1}{|c}{-} & 256 & 280 \\
                \midrule
                        \midrule
                        \diagbox{Cases}{\Op} & \#Mutants & \multicolumn{1}{|c}{\Ins} & \Del & \Name & \Attr \\
                        \midrule
                        Exhaustive & 25256 & \multicolumn{1}{|c}{25018} & 14 & 126 & 98 \\
                \midrule
                        Initial QCRMut\tnote{1} & 256 & \multicolumn{1}{|c}{256} & 0 & 0 & 0 \\
                        Apply 5\%\tnote{2} & 256 & \multicolumn{1}{|c}{217} & 13 & 13 & 13 \\
                Increase Mutants\tnote{3} & 280 & \multicolumn{1}{|c}{238} & 14 & 14 & 14 \\
        \midrule
                        Final split & 280 & \multicolumn{1}{|c}{238} & 14 & 14 & 14 \\
                \bottomrule
        \end{tabular}
    \begin{tablenotes}
        \item[1] Similarity principle, use the probabilities obtained in the exhaustive data.
        \item[2] Representativity principle, use the prefix \% if it does not reach.
        \item[3] Coverability principle, we increase the mutants to meet coverability as we do not have enough. However, we keep all principles.
    \end{tablenotes}
    \end{threeparttable}}
        \caption{QCRMut mutants study for BV Algorithm.}
        \label{tab:BVMutExample}
\end{table}

In practice, coverability is handled during mutant calculation, so no backtracking is required. Once the mutation data have been generated, the execution method is fixed, and the mutant budget and operator split are known, we can consider now the broader tool workflow and how these components fit together to create QCRMut.

\subsection{Mutation phase II, QCRMut}

\begin{figure*}
    \centering
    \scalebox{0.85}{
\begin{tikzpicture}[node distance=2cm]

\node (start) [startstop] {Receive Quantum Circuit};
\node (createMut) [process, right of=start, xshift=4cm, yshift=1.5cm] {Mutants generation};
\node (createInp) [process, below of=createMut, yshift=-1cm] {Input generation};
\node (Mut) [process, right of=createMut, xshift=3cm, yshift=-1.5cm] {Execution};
\node (save) [startstop, right of=Mut, xshift=2.1cm] {Save Results};

\draw [arrow] (start.east) -- (createMut.west);
\draw [arrow] (start.east) -- (createInp.west);
\draw [arrow] (createMut.east) -- (Mut.west);
\draw [arrow] (createInp.east) -- (Mut.west);
\draw [arrow] (Mut) -- (save);

\end{tikzpicture}}    \caption{Simplified Testing Workflow.}
    \label{fig:WorkflowMutTest1}
\end{figure*}

Fig.~\ref{fig:WorkflowQCRMut} illustrates the overall architecture of our tool, showing the QCRMut workflow and its main building blocks. Two design decisions, exhaustive generation and all-insertion-data handling, are represented explicitly. In contrast, the reduction strategy is not shown as a separate decision node; instead, it is passed as a parameter during mutant-data generation. This choice simplifies the diagram without obscuring the operational flow. We briefly clarify the following steps:

\begin{itemize}
    \item \textbf{Generate \bIns{} data at random}: We impose an upper bound on random generation for the \Ins{} operator to avoid potential infinite loops. Although unlikely, this safeguard ensures termination. If the bound is reached, the tool switches to generating all insertion data.

    \item \textbf{Mutant subset selection}: This step enforces coverability whenever possible by randomly selecting at least one mutant for each operator in \Op{}$^-$ at each index.

    \item \textbf{Generate all \bIns{} data with \breduction}: This is the only stage in which \reduction{} is highlighted explicitly. This is deliberate: exhaustive mutant generation is performed only with \reduction{} enabled, due to the computational overhead of generating and executing very large mutant sets. For BV, exhaustive generation yields 25256 mutants with \reduction{}, whereas without \reduction{} the number becomes prohibitively large, 75950 mutants for this small circuit.
\end{itemize}

In addition, QCRMut provides optional keyword arguments. Users may
choose whether to save mutants and specify the storage location, set a
seed to ensure reproducibility, or request an exact number of
mutants. If no number is provided, QCRMut estimates an appropriate
value based on coverage. If the requested number is below the minimum
required for coverability, QCRMut honours the request and selects
mutants at random, without enforcing coverability.

\subsection{Automated testing framework}

This is the final step of our tool development. Up to this point, we have created a mutant generator that aligns with our design principles. We take this one step further by integrating an automated testing process directly into the tool. At present, the automation is tailored to our mutant-generation workflow, but it is designed to be extensible.

To the best of our knowledge, there is no explicit testing framework that supports this kind of automation in our setting. QMutPy relies on the MutPy library for the testing phase; however, this approach is tightly coupled to a specific execution environment and does not readily extend to alternative testing approaches.

The framework follows the workflow in Fig.~\ref{fig:WorkflowMutTest1}, which outlines the overall testing process. Fig.~\ref{fig:WorkflowMutTest2} then details the execution flow for an individual mutant.

We divide the process into two classes. First, we collect the information required to perform mutation testing in quantum computing, including the original algorithm, mutant generator, input generator, test-circuit generator, execution method, and analysis function. These components are stored as attributes and exposed via class properties.

Second, we introduce a test class that instantiates the framework and supplies concrete parameters for the functions defined in the main class. This separation allows multiple tests with different configurations to reuse the same underlying structure.

\begin{figure*}
    \centering
    \scalebox{0.82}{
\begin{tikzpicture}[node distance=2cm]

\node (start) [startstop, text width=4cm] {Receive Mutant and Input list};
\node (createQC) [process, right of=start, xshift=4cm, text width=3.5cm] {Create Test QC for an Input};
\node (execute) [process, right of=createQC, xshift=3.5cm] {QC Execution};
\node (analysis) [process, below of=execute, yshift=-2cm] {Results Analysis};
\node (iskilled) [decision, below of=analysis, yshift=-2cm] {Mutant killed?};
\node (moreInp) [decision, left of=analysis, xshift=-3.5cm] {More inputs?};
\node (notkilled) [startstop, below of = start, yshift = -2cm] {Mutant Survived};
\node (killed) [startstop, below of = notkilled, yshift=-2cm] {Mutant Killed};

\draw [arrow] (start) -- (createQC);
\draw [arrow] (createQC) -- (execute);
\draw [arrow] (execute) -- (analysis);
\draw [arrow] (analysis) -- (iskilled);
\draw [arrow] (iskilled) -- node[anchor=south] {Yes} (killed);
\draw [arrow] (iskilled) -- node[anchor=north] {No} (moreInp);
\draw [arrow] (moreInp) -- node[anchor=south] {No} (notkilled);
\draw [arrow] (moreInp) -- node[anchor=east] {Yes} (createQC);

\end{tikzpicture}}
\caption{Execution Workflow for a Mutant.}
\label{fig:WorkflowMutTest2}
\end{figure*}

At present, our framework includes a few fixed functions. Specifically, the mutant generator is QCRMut, the execution environment is Qiskit's \lstinline|AerSimulator|, and the analysis method is \lstinline|chiSquare| to maintain compatibility with Muskit. The remaining functions must be provided by the user; however, the framework allows users to supply all necessary parameters for them. Some examples of different approaches are presented in next section. For instance, Table~\ref{tab:InputExample} shows different input locations, which are then implemented in the test circuit generator when instantiating inputs and measurements.

Before concluding this section, it is important to ensure that the statistical test employed produces reliable results. To this end, we calculate the number of executions required for each input to obtain statistically meaningful outcomes. These calculations are performed during the oracle computation, taking into account the lowest probabilities.

Because quantum measurements yield probabilistic outcomes, each circuit is executed for a number of shots to ensure the observed output distribution is statistically representative of the statevector. Therefore,  we must establish bounds for the number of shots. As a lower bound, we adopt 1024 shots, following the Qiskit default. The upper bound is determined from a target precision of $10^{-5}$ for the smallest probability. Although this may seem counter-intuitive, lower probabilities require more executions to guarantee at least 13 occurrences of each outcome, consistent with recommendations for the \lstinline|chiSquare| test. Under this precision, we set the upper bound to 1.3 million shots, ensuring that even very low-probability events are adequately sampled.


\section{Empirical experiments}
\label{S5:Exp}

The aim of this section is to present the experimental set-up and results needed to answer the proposed research questions. We conduct two main experiments. The first, the mutation experiment, provides insight into different mutation approaches and their scores, enabling comparison between random and exhaustive results. The second, the seed experiment, repeats the analysis for several random seeds to assess how the seed affects their score.

The tool, experimental results, data structures, and replication instructions are available in the provided \href{https://github.com/sinugarc/QCRMutPaper}{GitHub repository}\footnote{\url{hhttps://github.com/sinugarc/QCRMutPaper}}. All experiments were executed using Python 3.12, Qiskit 2.1, and Debian 12 on a machine with a 6-core/12-thread Intel(R) Xeon W-2235 CPU and 128~GB of RAM.

Before presenting the experiments, we introduce the quantum circuits for our experimental set-up.

\subsection{Quantum circuits data set}

One of our aims is to compare our approach with previous work; therefore, we seek to execute the same or similar quantum circuits. However, Qiskit Aqua, used by QMutPy~\cite{fortunato2022qmutpy}, has been deprecated in Qiskit stable version.

We have selected 21 quantum circuits, QCs, for our experiments. These are obtained from three different sources. First, the four proposed by Muskit~\cite{mendiluze2021muskit} in their original work:

\begin{itemize}
    \item \textbf{LittleBV}: Bernstein–Vazirani cryptography algorithm.
    \item \textbf{QRAM}: Quantum Random Access Memory.
    \item \textbf{CE}:
      Conditional Execution programme.
    \item \textbf{IQFT}: Inverse Quantum Fourier Transform.
\end{itemize}

Second, we include three additional oracle-input-based algorithms to demonstrate the functionality provided by our tool for this class of application:

\begin{itemize}
    \item \textbf{Simon}: Quantum routine from Simon’s algorithm.
    \item \textbf{DJ}: Deutsch–Jozsa algorithm.
    \item \textbf{BV}: Bernstein–Vazirani algorithm.
\end{itemize}

We note that both LittleBV and BV are instances of the Bernstein--Vazirani algorithm. Here, BV denotes the general algorithm, where the input and number of qubits can vary, whereas LittleBV is a fixed instance provided by Muskit~\cite{mendiluze2021muskit}.

\begin{table}
    \centering
    \resizebox{\columnwidth}{!}{%
    \begin{tabular}{|l|rr|rr|}
        \toprule
        \multirow{2}{*}{Quantum Circuit} & \multicolumn{2}{c|}{\# Gates} & \multicolumn{2}{c|}{\# Mutants} \\
        \cmidrule{2-5}
        & ------ & Mutable & QCRMut & Exhaustive \\
        \midrule
        LittleBV & 9 & 9 & 256 & 14796\\
        QRAM & 9 & 9 & 256 & 21765\\
        CE & 17 & 16 & 320 & 30866\\
        IQFT & 24 & 24 & 500 & 43881\\
        \midrule
        Simon & 13 & 12 & 256 & 50820\\
        DJ & 15 & 14 & 280 & 25256\\
        BV & 15 & 14 & 280 & 25256\\
        \midrule
        ghz & 6 & 6 & 256 & 13543\\
        graphstate & 12 & 12 & 256 & 20169\\
        wstate & 21 & 21 & 440 & 32408\\
        qft & 24 & 24 & 500 & 43881\\
        qpeexact & 27 & 27 & 560 & 44986\\
        qpeinexact & 28 & 28 & 580 & 46925\\
        vqe & 28 & 28 & 580 & 39369\\
        qftentangled & 30 & 30 & 620 & 53803\\
        qaoa & 30 & 30 & 620 & 44583\\
        ae & 41 & 31 & 840 & 68754\\
        portfoliovqe & 69 & 69 & 1680 & 108000\\
        realamprandom & 69 & 69 & 1680 & 108000\\
        su2random & 69 & 45 & 1680 & 107040\\
        twolocalrandom & 69 & 69 & 1680 & 108000\\
        \midrule
        \multicolumn{3}{|r|}{\# Total Mutants}& 14096 & 1052101 \\
        \bottomrule
    \end{tabular}%
    }
    \caption{Quantum Algorithms.}
    \label{tab:algorithms}
\end{table}

Finally, the remaining circuits are taken from MQT Benchmarks~\cite{quetschlich2023mqtbench}, which also served as the basis for the mutation benchmark used with Muskit~\cite{mendiluze2025quantumBench}. We include all scalable benchmarks, instantiating them on six qubits, following Muskit~\cite{mendiluze2021muskit}. We exclude the random circuit, preferring circuits that solve or represent a defined problem, and we omit circuits with more than 100 gates due to computational resource limitations.

Table~\ref{tab:algorithms} summarises the selected circuits and reports, for each circuit, the number of mutable gates and the number of mutants generated by our tool using both the gate-coverage and exhaustive methods.

\begin{figure*}
    \centering
    \includegraphics[width=0.97\textwidth]{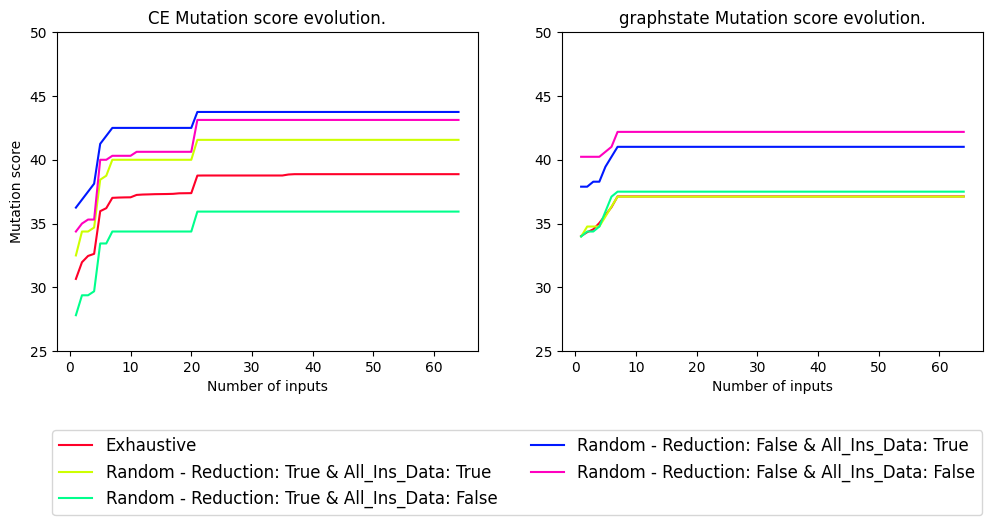}
    \caption{Experiment I, Mutation Score.}
    \label{FIG:MutationScore}
\end{figure*}

\subsection{Mutation experiment}

We now turn to the main experiment of this paper. Our objective is to execute all selected quantum circuits using the exhaustive method and to evaluate the full space of configurations in our random approach (e.g., with and without \reduction{} and with and without \allins{}). Recall that \allins{} denotes generating all insertion data and then selecting the desired subset, whereas \reduction{} controls whether we apply our reduction procedure based on the insertion position in the graphical representation.

The experimental set-up builds on our automated testing framework, with input configuration and result analysis following the structure of the Muskit experiments~\cite{mendiluze2021muskit}.

For result analysis, we distinguish only between \emph{killed} and \emph{non-killed} mutants. We use \lstinline|chiSquare| from SciPy library with significance level $p = 0.05$. In our configuration, the test assumptions are satisfied because the number of circuit executions is determined by the lowest probability in the oracle, ensuring the required minimum observed frequencies.

An exception occurs for the algorithms we introduce (Simon, DJ, and BV), where the input is instantiated within the corresponding input-holder. Table~\ref{tab:InputExample} illustrates this.

\begin{table}
  \begin{center}
\newsavebox{\cirLBV}
\newsavebox{\cirBV}
\savebox{\cirLBV}{
  \begin{quantikz}
    & \gate[6,disable auto
height]{\verticaltext{INPUT}} & \qw & \ctrl{3} & \qw & \qw & \qw & \\
    & &\qw & \qw & \ctrl{3} & \qw & \qw &\\
    & &\qw & \qw & \qw & \ctrl{3} & \qw &\\
    & &\gate{H} & \gate{Z} & \qw & \qw & \gate{H} &\\
    & &\gate{H} & \qw & \gate{Z}& \qw & \gate{H} &\\
    & &\gate{H} & \qw & \qw & \gate{Z}& \gate{H} &
  \end{quantikz}
}
\savebox{\cirBV}{
  \begin{quantikz}
    &\gate{H} & \qw & \gate[7,disable auto
height]{\verticaltext{INPUT}}  & \gate{H} &\\
    &\gate{H} & \qw &  &\gate{H} &\\
    &\gate{H} & \qw &  &\gate{H} &\\
    &\gate{H} & \qw &  &\gate{H} &\\
    &\gate{H} & \qw &  &\gate{H} &\\
    &\gate{H} & \qw &  &\gate{H} &\\
    &\gate{H} &  \gate{Z} & &\gate{H} &
  \end{quantikz}
}
\resizebox{0.95\columnwidth}{!}{%
  \begin{tabular}{|l|c|c|}
    \toprule
      &  LittleBV - Initialisation & BV - Oracle \\
    \midrule
    QC &
    \usebox\cirLBV&
    \usebox\cirBV
    \\
    \bottomrule
  \end{tabular}%
}
\end{center}
\caption{Input position examples.}
\label{tab:InputExample}
\end{table}

Table~\ref{tab:ExpI} reports the results. For each algorithm, we also generate a figure to compare the evolution of mutation score across configurations. For brevity, we show one example in Fig.~\ref{FIG:MutationScore}; the full set of figures is provided in the Appendix.

We now introduce the metrics and tests used to analyse these results, providing the first answers to RQ2 and RQ3.
First, for RQ2, we aim to identify the random configuration that most closely approximates the exhaustive results. We use the root mean square error, RMSE, between exhaustive mutation scores and those obtained from each random configuration. The full analysis is reported in Table~\ref{tab:MutTestMAE}.

\begin{table}
    \centering
    \resizebox{0.95\columnwidth}{!}{%
    \begin{tabular}{|c|cccc|}
    \toprule
         $(\reduction, \allins)$ & (T, T) & (T, F) & (F, T) & (F, F) \\
         \midrule
         RMSE & 3.31 & 1.85 & 4.60 & 6.40 \\
         \bottomrule
    \end{tabular}%
    }
    \caption{Mutation Test - RMSE Results.}
    \label{tab:MutTestMAE}
\end{table}

In response to RQ2, the closest configuration by RMSE ($1.85$) uses \reduction{} but not \allins{}. We therefore select this as the default setting in QCRMut. Second, for RQ3, we assess whether our selected random approach yields mutation scores that are statistically indistinguishable from the exhaustive results. We employ two non-parametric measures that do not rely on distributional assumptions and reduce the risk of misleading statistical interpretations: the Wilcoxon signed-rank test and Cliff’s $\delta$ effect size.

For this analysis, we consider only mutation scores below 100\%. Scores of 100\% can mislead the test, due to any non-empty subset also yielding 100\%. The results are as follows:

\vspace{-5pt}
\begin{itemize}
    \item[] Wilcoxon test, $p-value = 0.1475$.
    \item[] Cliff's $\delta$ $= -0.074$, effect size: negligible.
\end{itemize}

\vspace{-7pt}
We therefore conclude that no statistically significant differences are detected ($p$-value $> 0.05$) and the effect size is negligible. These results support the validity of our proposed random approach.

\begin{landscape}

\begin{table}
    \centering
    \resizebox{\columnwidth}{!}{%
    \begin{tabular}{|l||r|rrrr||r|rrrr|}
        \toprule
        \multirow{2}{*}{Quantum Circuit} & \multicolumn{5}{c||}{Mutation Score (\%) - (\reduction, \allins)} & \multicolumn{5}{c|}{Execution time (s) - (\reduction, \allins)}\\
        \cmidrule{2-11}
        & Exhaustive & (T, T) & (T, F) & (F, T) & (F, F) & Exhaustive & (T, T) & (T, F) & (F, T) & (F, F) \\
        \midrule
        LittleBV & 33.13 & 32.42 & 35.94 & 38.67 & 41.80 & 16546.94 & 294.35 & 289.95 & 279.25 & 276.82 \\
        QRAM & 53.58 & 59.77 & 55.08 & 47.66 & 51.56 & 34879.57 & 374.71 & 412.93 & 489.34 & 446.58 \\
        CE & 38.87 & 41.56 & 35.94 & 43.75 & 43.12 & 32279.98 & 339.66 & 371.17 & 324.98 & 320.56 \\
        IQFT & 57.88 & 60.00 & 59.80 & 60.80 & 62.60 & 33778.79 & 366.19 & 368.72 & 353.30 & 336.11 \\
        \midrule
        Simon & 64.06 & 67.19 & 66.02 & 70.70 & 71.48 & 34526.58 & 178.81 & 170.37 & 152.31 & 155.08 \\
        DJ & 72.96 & 78.57 & 75.00 & 77.50 & 82.86 & 24315.01 & 244.62 & 281.39 & 249.12 & 218.11 \\
        BV & 74.00 & 78.21 & 75.00 & 78.21 & 84.64 & 11971.63 & 120.24 & 142.83 & 120.86 & 90.34 \\
        \midrule
        ghz & 73.54 & 70.70 & 75.78 & 79.30 & 79.69 & 6374.43 & 161.35 & 131.04 & 105.41 & 104.30 \\
        graphstate & 37.13 & 37.11 & 37.50 & 41.02 & 42.19 & 21247.97 & 291.29 & 297.05 & 286.11 & 280.67 \\
        wstate & 100.00 & 100.00 & 100.00 & 100.00 & 100.00 & 3847.21 & 13.13 & 12.69 & 12.95 & 13.31 \\
        qft & 100.00 & 100.00 & 100.00 & 100.00 & 100.00 & 1176.84 & 13.97 & 14.17 & 13.50 & 13.64 \\
        qpeexact & 99.89 & 100.00 & 99.82 & 99.82 & 100.00 & 4736.72 & 28.81 & 31.73 & 29.51 & 31.87 \\
        qpeinexact & 100.00 & 100.00 & 100.00 & 100.00 & 100.00 & 7893.84 & 23.12 & 25.01 & 25.03 & 25.72 \\
        vqe & 100.00 & 100.00 & 100.00 & 100.00 & 100.00 & 2282.88 & 17.96 & 18.24 & 17.68 & 17.69 \\
        qftentangled & 87.98 & 89.68 & 86.94 & 90.00 & 89.03 & 52078.00 & 288.86 & 360.50 & 277.12 & 308.67 \\
        qaoa & 100.00 & 100.00 & 100.00 & 100.00 & 100.00 & 17737.43 & 85.42 & 87.98 & 87.37 & 90.20 \\
        ae & 100.00 & 100.00 & 100.00 & 100.00 & 100.00 & 5811.62 & 41.43 & 43.20 & 41.05 & 43.97 \\
        portfoliovqe & 100.00 & 100.00 & 100.00 & 100.00 & 100.00 & 24061.16 & 128.92 & 128.89 & 128.93 & 128.91 \\
        realamprandom & 100.00 & 100.00 & 100.00 & 100.00 & 100.00 & 24134.71 & 131.08 & 134.20 & 132.35 & 132.66 \\
        su2random & 100.00 & 100.00 & 100.00 & 100.00 & 100.00 & 20717.21 & 92.75 & 96.29 & 93.32 & 93.97 \\
        twolocalrandom & 100.00 & 100.00 & 100.00 & 100.00 & 100.00 & 25195.93 & 131.18 & 134.61 & 132.56 & 132.52 \\
        \bottomrule
    \end{tabular}%
    }
  \caption{Experiment I, All.}
  \label{tab:ExpI}
\end{table}
\end{landscape}

Before concluding this section, we consider Table~\ref{tab:ExpI} from a computational perspective. In addition to mutation score, we report the execution time required to simulate each test. A key motivation for this work is to reduce computational cost, given the resources required for circuit execution. The concept of \emph{do fewer} was established in the literature by Usaola et al.~\cite{usaola2010mutation} and Pizzoleto et al.~\cite{pizzoleto2019systematic}. QCRMut follows this principle while still obtaining comparable results.

As shown in Table~\ref{tab:algorithms}, the exhaustive method executes at least one million mutants. We emphasise \emph{at least} because this lower bound would occur only if every mutant were killed by the first input, which is not the case in practice. In contrast, under QCRMut, we execute just over 14000 mutants.

\subsection{Seed Experiment}

We have shown that our random approach provides a close approximation to the exhaustive results. However, since mutant generation depends on a random seed (used primarily for reproducibility), it is reasonable to ask whether the results in the previous section depend on the chosen seed. This section shows that they do not.

To address this, we conduct a \emph{seed test} by repeating the experiments using different random seeds. Specifically, we consider the natural numbers 1–10, together with an additional set of ten seeds drawn uniformly from the interval $[0,1]$ using Python’s \lstinline|random| library. For ease of notation and comparison, these seeds are ordered prior to analysis.

We restrict this analysis to circuits whose exhaustive mutation scores are below 100\%, as only these cases can reveal differences across seeds. For each selected circuit, we execute progressively larger numbers of mutants to track mutation score evolution. This allows us to study each algorithm up to the number of mutants provided by QCRMut to ensure coverability, and to analyse its behaviour beyond that point.

Fig.~\ref{FIG:ghzSeedMutScore} illustrates the experiment for \emph{ghz}; the remaining figures are provided in the Appendix. The figure shows how mutation score evolves as more mutants are executed. For larger numbers of mutants, the choice of seed has no material effect: trajectories remain within a narrow range. As expected, when only a small number of mutants is executed, mutation score becomes sensitive to the seed, with differences of up to 50\%. This arises because the seed determines the specific mutant subset. For example, if the exhaustive mutation score is 70\% (based on more than 10000 mutants), selecting only 25 or 50 mutants may yield substantially different subsets and therefore different scores. This observation motivates our emphasis on coverability in the tool. We use supporting graph lines indicating the number of mutants used, together with the observed minimum and maximum. This facilitates a graphical assessment of the stability of seed-based results with respect to the number of mutants configured in the tool.

Overall, the graphical results indicate that the generation seed does not compromise the reliability of mutation score: for the prescribed number of mutants (and beyond), the score stabilises regardless of the seed. We nevertheless analyse seed sensitivity more precisely using the same statistics as in the previous experiment: RMSE, Wilcoxon, and Cliff’s $\delta$.

Firstly, for each algorithm we compute the RMSE between the mutation scores obtained across all seeds (using QCRMut expected number of mutants) and the corresponding exhaustive mutation score. Table~\ref{tab:ghzSeed} gives an example for \emph{ghz}, evaluated using its expected 256 mutants. The results of this initial analysis are reported in Table~\ref{tab:SeedExpFirstAnalysis}.

\begin{figure*}
    \centering
    \includegraphics[width=0.85\textwidth]{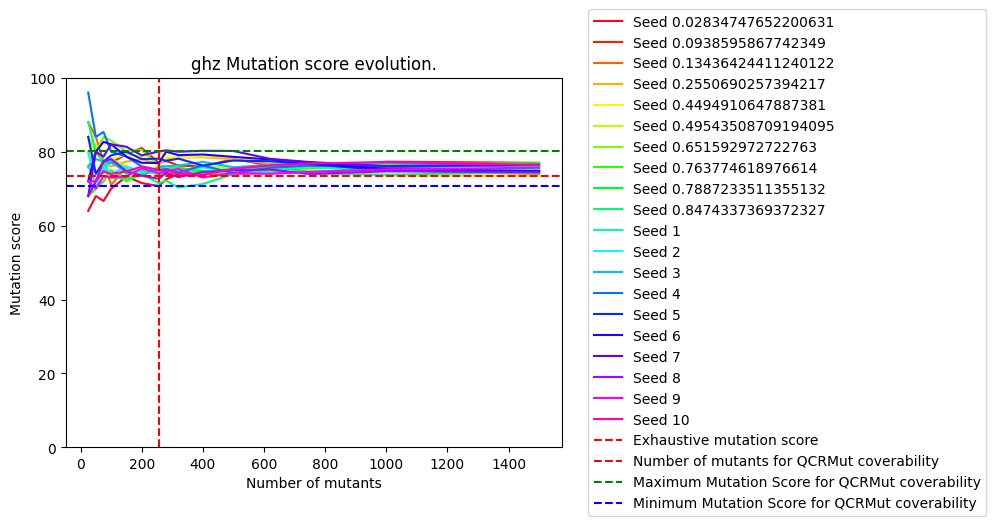}
    \caption{Seed Experiment, ghz Mutation Score evolution.}
    \label{FIG:ghzSeedMutScore}
\end{figure*}

Let us clarify the notation used in Tables~\ref{tab:ghzSeed} and \ref{tab:SeedExpI}. The prefix R denotes a randomly generated seed, and the index refers to its position within the ordered list. For example, R1 $= 0.028\ldots$ and R10 $= 0.847\ldots$ . In contrast, the natural numbers denote themselves as seeds.

\begin{table}[H]
    \centering
    \resizebox{\columnwidth}{!}{%
    \begin{tabular}{|c|cccccccccc|}
        \toprule
         Seed & R1 & R2 & R3 & R4 & R5 & R6 & R7 & R8 & R9 & R10 \\
         \midrule
         M$_{Score}$ & 70.70 & 76.95 & 75.00 & 76.56 & 80.08 & 76.17 & 76.17 & 75.39 & 73.83 & 71.48 \\
         \midrule
         \midrule
         Seed & 1 & 2 & 3 & 4 & 5 & 6 & 7 & 8 & 9 & 10 \\
         \midrule
         M$_{Score}$ & 75.78 & 74.22 & 75.78 & 75.00 & 78.12 & 76.95 & 80.08 & 72.66 & 74.22 & 75.00 \\
         \midrule
         \midrule
         \multicolumn{3}{|c|}{Exh Mut Score} & \multicolumn{2}{c|}{73.54}&\multicolumn{4}{c|}{RMSE} & \multicolumn{2}{c|}{3.06}\\
         \bottomrule
    \end{tabular}}
    \caption{GHZ algorithm mutation score results for 256 mutants and analysis.}
    \label{tab:ghzSeed}
\end{table}

\begin{table}[H]
    \centering
    \resizebox{\columnwidth}{!}{%
    \begin{tabular}{|c|ccccc|}
        \toprule
        Algorithm & LittleBV & QRAM & CE & IQFT & Simon \\
        \midrule
        RMSE & 6.36 & 4.14 & 3.41 & 2.96 & 2.10 \\
        \midrule
        \midrule
        Algorithm & BV & ghz & graphstate & qftentangled & - \\
        \midrule
        RMSE & 4.62 & 3.06 & 1.68 & 1.21 & - \\
        \bottomrule
    \end{tabular}}
    \caption{Seed Experiment; RMSE results.}
    \label{tab:SeedExpFirstAnalysis}
\end{table}

\begin{table}
    \resizebox{\columnwidth}{!}{%
    \begin{threeparttable}
    \begin{tabular}{|c||c|cccccccccc|}
        \toprule
        \diagbox{\tnote{2}}{\tnote{1}} & Exh & R1 & R2 & R3 & R4 & R5 & R6 & R7 & R8 & R9 & R10 \\
        \midrule
        \midrule
        Exh & - & \textbf{0.039} & 0.129 & 0.098 & \textbf{0.039} & \textbf{0.027} & 0.250 & \textbf{0.020} & 0.570 & 0.652 & 0.359 \\
        \midrule
        R1 & neg & - & 0.734 & 0.779 & 0.779 & 0.426 & 0.734 & 0.910 & 0.496 & 0.779 & 1.000 \\
        R2 & neg & neg & - & 0.652 & 0.426 & \textbf{0.025} & 0.910 & 0.398 & 0.820 & 0.575 & 0.779 \\
        R3 & neg & neg & neg & - & 0.496 & 0.161 & 0.570 & 0.889 & 0.570 & 0.570 & 1.000 \\
        R4 & \textbf{sma} & neg & neg & neg & - & 0.570 & 0.570 & 0.910 & 0.426 & 0.496 & 1.000 \\
        R5 & \textbf{sma} & neg & \textbf{sma} & neg & neg & - & 0.098 & 0.250 & 0.164 & 0.203 & 0.570 \\
        R6 & neg & neg & neg & neg & neg & neg & - & 0.674 & 0.779 & 0.652 & 0.652 \\
        R7 & neg & neg & neg & neg & neg & neg & neg & - & 0.250 & 0.496 & 0.833 \\
        R8 & neg & neg & neg & neg & neg & \textbf{sma} & neg & neg & - & 0.820 & 1.000 \\
        R9 & neg & neg & neg & neg & neg & neg & neg & neg & neg & - & 1.000 \\
        R10 & neg & neg & neg & neg & neg & neg & neg & neg & neg & neg & - \\
        \midrule
        1 & neg & neg & neg & neg & neg & \textbf{sma} & neg & neg & neg & neg & neg \\
        2 & neg & neg & neg & neg & neg & \textbf{sma} & neg & neg & neg & neg & neg \\
        3 & neg & neg & neg & neg & neg & neg & neg & neg & \textbf{sma} & neg & neg \\
        4 & neg & neg & neg & neg & neg & \textbf{sma} & neg & neg & neg & neg & neg \\
        5 & \textbf{sma} & neg & neg & neg & neg & neg & neg & neg & neg & neg & neg \\
        6 & neg & neg & neg & neg & neg & neg & neg & neg & neg & neg & neg \\
        7 & \textbf{sma} & neg & neg & neg & neg & neg & neg & neg & \textbf{sma} & neg & neg \\
        8 & neg & neg & neg & neg & neg & neg & neg & neg & neg & neg & neg \\
        9 & \textbf{sma} & neg & neg & neg & neg & neg & neg & neg & neg & neg & neg \\
        10 & neg & neg & neg & neg & neg & neg & neg & neg & neg & neg & neg \\
        \toprule
        \toprule
        \diagbox{\tnote{2}}{\tnote{1}} & - &1 & 2 & 3 & 4 & 5 & 6 & 7 & 8 & 9 & 10 \\
        \midrule
        \midrule
        Exh & - & 0.250 & 0.496 & 0.164 & 0.359 & 0.055 & 0.250 & \textbf{0.027} & 0.820 & 0.074 & 0.164 \\
        \midrule
        R1 & - & 0.250 & 0.652 & 0.734 & 0.734 & 0.652 & 0.652 & 0.250 & 0.652 & 0.734 & 0.652 \\
        R2 & - & 0.496 & 0.734 & 0.575 & 0.820 & 0.301 & 0.866 & 0.164 & 0.263 & 0.910 & 0.734 \\
        R3 & - & 0.207 & 0.262 & 0.359 & 0.484 & 0.164 & 0.652 & 0.203 & 0.426 & 0.400 & 0.484 \\
        R4 & - & 0.164 & 0.203 & 0.944 & 0.092 & 0.652 & 0.910 & 0.301 & 0.203 & 1.000 & 0.092 \\
        R5 & - & 0.055 & 0.055 & 0.250 & 0.074 & 0.734 & 0.359 & 0.889 & \textbf{0.039} & 0.426 & \textbf{0.036} \\
        R6 & - & 0.440 & 0.359 & 0.426 & 0.574 & 0.426 & 0.820 & 0.098 & 0.496 & 0.652 & 0.426 \\
        R7 & - & 0.301 & 0.237 & 0.674 & 0.496 & 0.496 & 1.000 & 0.164 & 0.441 & 0.674 & 0.652 \\
        R8 & - & 0.779 & 1.000 & 0.164 & 0.726 & 0.250 & 0.164 & \textbf{0.036} & 1.000 & 0.426 & 0.734 \\
        R9 & - & 0.820 & 0.496 & 0.359 & 0.734 & 0.203 & 0.301 & 0.055 & 0.570 & 0.301 & 0.779 \\
        R10 & - & 0.820 & 0.575 & 0.820 & 0.674 & 0.496 & 0.570 & 0.301 & 0.441 & 0.570 & 0.734 \\
        \midrule
        1 & - & - & 0.910 & 0.161 & 0.652 & 0.129 & 0.164 & 0.055 & 0.820 & 0.203 & 0.484 \\
        2 & - & neg & - & 0.123 & 0.441 & 0.074 & \textbf{0.049} & \textbf{0.017} & 1.000 & 0.128 & 0.499 \\
        3 & - & \textbf{sma} & neg & - & 0.068 & 0.652 & 0.944 & 0.301 & 0.203 & 1.000 & 0.150 \\
        4 & - & neg & neg & neg & - & 0.129 & 0.203 & \textbf{0.012} & 0.910 & 0.207 & 0.446 \\
        5 & - & neg & \textbf{sma} & neg & neg & - & 0.652 & 0.734 & 0.129 & 0.401 & 0.250 \\
        6 & - & neg & neg & neg & neg & neg & - & 0.570 & 0.401 & 1.000 & 0.570 \\
        7 & - & \textbf{sma} & \textbf{sma} & neg & \textbf{sma} & neg & neg & - & 0.098 & 0.575 & \textbf{0.036} \\
        8 & - & neg & neg & neg & neg & neg & neg & neg & - & 0.161 & 0.570 \\
        9 & - & neg & neg & neg & neg & neg & neg & neg & neg & - & 0.301 \\
        10 & - & neg & neg & neg & neg & neg & neg & neg & neg & neg & - \\
        \bottomrule
    \end{tabular}
    \begin{tablenotes}
        \item[1] Wilcoxon $p-value$. \item[2] Cliff’s $\delta$ Effect-Size. Legend: neg := negligible; sma := small.
    \end{tablenotes}
    \end{threeparttable}}
    \caption{Seed Experiment Analysis.}
    \label{tab:SeedExpI}
\end{table}

Finally, we compute the remaining statistics for all pairwise seed combinations, analogous to the previous experiment. The results are presented in Table~\ref{tab:SeedExpI}, with $p$-values below $0.05$ shown in bold and effect sizes indicated when the result is not negligible.

There are 210 pairwise combinations. Of these, 13 yield a $p$-value below $0.05$, suggesting statistically significant differences. However, we also consider effect size and treat negligible effect sizes as acceptable. Under this criterion, we reduce the set to 7 combinations (Table~\ref{tab:SeedAnalysis}). The resulting RMSE values range from 2.54 to 4.34, which we consider an acceptable variation given that we reduce the number of mutants by more than 90\%.

We can now answer \textbf{RQ4}. The seed does not significantly affect results in almost 95\% of cases. Among the remaining cases, roughly half have negligible effect size; the seven cases with small effect size are shown in Table~\ref{tab:SeedAnalysis}. We consider the corresponding RMSE variation acceptable in light of the computational savings, and note that these cases represent only 3.33\% of all combinations.

To conclude this section, we summarise the results so far. We have answered \textbf{RQ2}, \textbf{RQ3}, and \textbf{RQ4}, which enables us to address \textbf{RQ1}. We identify a tool configuration for which no statistically significant differences from the exhaustive procedure are detected. We also show that the seed does not significantly affect the results produced by the mutant generator. We can therefore conclude that our randomly generated mutants are representative of those generated exhaustively.

\begin{table}[H]
    \centering
    \resizebox{\columnwidth}{!}{%
    \begin{tabular}{|c|c||c|c||c|}
        \toprule
         Seed 1 & Seed 2 & Wilcoxon $p-value$ & Cliff’s $\delta$ Effect-Size & RMSE \\
         \midrule
         \midrule
         Exh & R4 & $0.039$ & \multirow{7}{*}{small} & $3.51$\\
         Exh & R5 & $0.027$ &  & $4.34$ \\
         Exh & 7 & $0.027$ &  & $4.29$ \\
         R2 & R5 & $0.025$ &  & $2.54$ \\
         R8 & 7 & $0.036$ &  & $3.39$ \\
         2 & 7 & $0.017$ &  & $3.38$ \\
         4 & 7 & $0.012$ &  & $2.92$ \\
         \bottomrule
    \end{tabular}}
    \caption{Seed combination analysis.}
    \label{tab:SeedAnalysis}
\end{table}

\subsection{Comparison with previous tools}

Having shown that our tool can generate a representative subset of mutants whose behaviour is comparable to the exhaustive approach, we next compare with existing tools. We introduce each tool individually and analyse it separately to provide a clearer, more structured comparison.

\textbf{Muskit~\cite{mendiluze2021muskit}}.
  Mendiluze et
    al. introduced Muskit as the first automated mutation testing tool
    for quantum programs. The tool operates at the source-code level
    and follows an exhaustive strategy, generating all possible
    mutations at each identified gap within the quantum circuit. The
    resulting mutants are then executed and assessed using a provided
    oracle to determine whether they have been killed. The authors
    evaluate Muskit on four quantum algorithms, all of which are
    included in our experimental setup. Their reported results are
    summarised in Table~\ref{tab:MuskitResults} alongside our results,
    to facilitate ease of comparison.

\begin{table}[H]
    \centering
    \begin{tabular}{|c||r|r|}
        \toprule
        \multirow{2}{*}{Algorithm} & \multicolumn{2}{c|}{\% Mutation Score}  \\
        & Muskit & QCRMut \\
        \midrule
        IQFT & 100.00 & 57.88 \\
        QRAM & 93.93 & 53.58 \\
        LittleBV & 77.35 & 33.13 \\
        CE & 100.00 & 38.87 \\
        \bottomrule
    \end{tabular}
    \caption{Muskit~\cite{mendiluze2021muskit} vs QCRMut results.}
    \label{tab:MuskitResults}
\end{table}

Due to the large discrepancies between the two approaches when using the same original code, we decided that further investigation was required.

We begin with IQFT. Muskit executes this algorithm using only 100 repetitions, which can cause issues when applying the chosen statistical test. The \lstinline|chiSquare| test requires a minimum number of observations per outcome to yield meaningful results. This is particularly problematic for IQFT because, for each input, the algorithm may produce any of 64 possible outcomes. With only 100 repetitions, many outcomes are observed too rarely, making the statistical analysis unreliable; consequently, we cannot perform a fair comparison for IQFT.

In contrast, LittleBV and CE do not suffer from the same limitation. For these algorithms, the number of possible outcomes per input is at most four, and these outcomes are equiprobable. Under these conditions, 100 repetitions provide sufficient data for each outcome, allowing the \lstinline|chiSquare| test to be applied effectively.

For LittleBV, we deliberately identified a surviving mutant in our framework that is classified as killed by Muskit. The mutant \emph{Add4} satisfies this condition. This mutant applies the insertion operator by adding a Z gate on qubit 0 at the beginning of the circuit. Muskit’s oracle specifies that, for an input $n \in \mathbb{N}$, the expected output should be $n$. However, the measurement is restricted to three qubits, which limits the observable outcomes to values in the range $[0,7]$ when interpreted in binary form. Consequently, for any input greater than 7, the oracle condition can never be satisfied, and the mutant is invariably classified as killed.

An additional issue observed in these results concerns the use of the \lstinline|chiSquare| test to compare exact equalities. For example, for input 0, the observed result is also 0, which should satisfy the oracle. Nevertheless, the mutant is reported as killed because the resulting p-value is NaN. This highlights a further inconsistency in the application of the statistical test.

Given these discrepancies, we conclude that a fair and meaningful comparison with Muskit cannot be achieved for these cases. We therefore proceed with the evaluation of the next available tool in the state of the art.

\textbf{QMutPy~\cite{fortunato2022mutation,fortunato2022qmutpy}}.
    Fortunato et al. presented a quantum mutation testing tool
    developed as an extension of MutPy. Their approach employs both
    classical and quantum mutation operators and relies on MutPy’s
    existing structure for the mutation and execution
    processes. Assertions derived from the non-mutated quantum circuit
    are used to determine whether a mutant has been killed. Although
    the authors evaluate their tool on 24 algorithms, quantum mutation
    operators are applied to only 11 of them. The resulting outcomes
    are reported in Table~\ref{tab:QMutPyResults}. However, their metric differs slightly from ours; therefore, we include an additional column computing the mutation score using our formula, defined as the proportion of mutants killed over mutants fully executed, excluding still-born mutants.

    \vspace{10pt}

\begin{table}[H]
    \centering
    \resizebox{\columnwidth}{!}{%
    \begin{threeparttable}
        \begin{tabular}{|c||c|c||r|r||r|}
            \toprule
            \multirow{2}{*}{Algorithm} &\multicolumn{2}{c||}{QMutPy} & \multicolumn{3}{c|}{\% Mutation Score}  \\
             & \# Mutants\tnote{1} & \# Killed & QMutPy & QMutPy\tnote{2} & QCRMut \\
            \midrule
            \midrule
            bernstein\_vazirani & 93 & 74 & 91.32 & 79.57 & 75.00 \\
            deutsch\_jozsa & 93 & 66 & 87.68 & 70.97 & 75.00 \\
            grover & 93 & 17 & 50.32 & 18.28 & - \\
            grover\_optimizer & 52 (50) & 2 & 25.00 & 100.00 & - \\
            hhl & 2 (1) & 1 & 100.00 & 100.00 & - \\
            iqpe & 105 (4) & 82 & 90.56 & 81.19 & - \\
            qsvm & 1 & 1 & 100.00 & 100.00 & - \\
            shor & 207 (7) & 50 & 53.34 & 25.00 & - \\
            simon & 47 & 32 & 86.36 & 68.09 & 66.02 \\
            vqc & 2 (1) & 1 & 0.00 & 100.00 & - \\
            vqe & 1 (1) & - & - & - & 100.00 \\
            \bottomrule
        \end{tabular}
        \begin{tablenotes}
            \item[1] Mutants not exercised by the test suite or timeout on execution. \item[2] Using our formula with their data. We do not consider the not exercised or timeout mutants as part of the full set, this will always improve the score.
        \end{tablenotes}
    \end{threeparttable}}
    \caption{QMutPy~\cite{fortunato2022mutation,fortunato2022qmutpy} vs QCRMut results.}
    \label{tab:QMutPyResults}
\end{table}

Out of the 11 algorithms to which QMutPy applies quantum mutation operators, we execute four: BV, DJ, Simon, and vqe. For the first three, our results are consistent with QMutPy, despite generating substantially more mutants. QMutPy produces between 50 and 100 mutants per algorithm, whereas our random approach generates between 250 and 300 mutants to ensure coverability, and our exhaustive approach produces more than 25000 mutants. Direct comparisons of coverability are not possible, since our approach moves from line-of-code coverage to gate-level coverage.

For the remaining algorithm, vqe, QMutPy is only able to generate a single mutant, which subsequently times out during execution. In contrast, our approach produces 580 mutants using the random strategy and nearly 40000 mutants using the exhaustive strategy. These discrepancies are likely attributable to differences in the mutation operators and the mutation process: QMutPy relies on Python abstract syntax tree and the MutPy framework, whereas our approach operates directly on the internal circuit structure.\\

Finally, we note that for some algorithms to which QMutPy is unable to apply quantum mutation operators, such as qaoa and qpe, our tool successfully generates mutants and achieves near-perfect mutation scores.

We can therefore answer \textbf{RQ5}. Whenever a direct comparison with existing tools is possible, our results are consistent with those previously reported, despite not relying on exhaustive generation. At the same time, our approach supports a broader set of gates and mutation operators, enabling a wider and more expressive mutant set that better captures realistic developer errors in quantum circuits. Moreover, QCRMut can generate and analyse mutants in scenarios where other approaches fail. Consequently, QCRMut not only meets current state-of-the-art standards but also overcomes key limitations of prior tools.

\vspace{10pt}
\subsection{Threats to validity}

\vspace{10pt}
\begin{itemize}
\item \textbf{Internal validity}.
  Reliability is affected by the tool's randomised strategies and the
  use of explicit seeds. We therefore evaluate variability across
  different seed values. Our analysis shows no significant differences
  in most cases. In the few instances where minor differences are
  observed, they are attributable to the computational savings obtained
  by executing fewer mutants. Although we address this in the study, we
  explicitly acknowledge it as an important concern.
  
  Another threat involves the identification of equivalent
  mutants. While full behavioural equivalence is often computationally
  undecidable, QCRMut utilizes graphical equivalence to prune
  redundant mutants. Although this is a significant advancement over
  existing quantum mutation tools that ignore circuit structure, it
  may not capture all identical behaviours. Consequently, the reported
  mutation scores might be influenced by remaining undetected
  duplicates, a limitation we aim to address in future work through
  more exhaustive equivalence-checking techniques.

\newpage
\item \textbf{External validity}.
  We aim to include as many representative quantum algorithms as
  possible. To this end, we select an established
  benchmark~\cite{quetschlich2023mqtbench} and apply our approach to
  these algorithms, complemented by an extension involving
  oracle-based algorithms. Nevertheless, our evaluation is limited to
  algorithms operating on six qubits, which may threaten external
  validity. Although we do not expect the number of qubits to
  introduce fundamental differences in the observed behaviour, this
  warrants further investigation. Our choice is consistent with
  Muskit~\cite{mendiluze2021muskit}, which also restricts its
  evaluation to six-qubit circuits.

\item \textbf{Construct validity}. Several construct-validity
  concerns remain open, mainly relating to design decisions
  adopted to satisfy our guiding principles. Examples include
  setting a minimum representativity bound of $5\%$, restricting
  parameters to seven discrete values (even though values over
  $(0,\pi)$ yield distinct mutants), and using a reduction scheme
  for exhaustive generation. Although we have not analysed these
  decisions exhaustively, they appear reasonable in light of our
  experimental results. We nevertheless identify systematic
  analysis of these parameters as future work, to better
  understand their impact on variability and effectiveness.

\item \textbf{Conclusion validity.}
  It arises from the challenges discussed above when comparing our
  approach with existing tools. These challenges include
  inconsistencies in reported results, differences in gate sets,
  and variation across Qiskit versions. Nonetheless, this work
  advances the state of the art by providing a tool compatible
  with stable versions of Qiskit that can be readily used and
  extended by the community.

\end{itemize}


\section{Conclusion}
\label{S6:Conclusion}
Through this paper, we introduce QCRMut as a novel alternative approach for generating mutants of quantum circuits, designed to operate with stable versions of Qiskit. The tool offers flexibility to maintain a desired structure for algorithm robustness. Our method provides an efficient means of generating random mutants that are correct by construction, producing results comparable to those produced by exhaustive mutant generators.

In contrast to existing mutation testing approaches, both classical and quantum, which predominantly operate at the level of lines of code, LOC, our approach is quantum-oriented and focuses instead on the number of gates within an algorithm. This perspective affords users greater flexibility, and is further strengthened by the integration of an automated testing component within the framework.

Our proposed execution aligns with previous works by Mendiluze et al. \cite{mendiluze2021muskit}, which require the use of an oracle. Alternatively, Fortunato et al. \cite{fortunato2022qmutpy} employ an assertion-based approach, utilising a non-mutated version of the algorithm to derive results. Through a comparative analysis, we have shown that, where fair comparisons are possible, QCRMut produces results consistent with those reported by existing tools, while also supporting a broader range of gate sets, and algorithms. Moreover, QCRMut is able to generate mutants in scenarios where previous approaches are unable to do.

A further contribution of this work lies in its explicit treatment of statistical validity. We highlighted limitations in the application of statistical tests across the state of the art and addressed these concerns by dynamically determining the number of executions required based on the lowest admissible probability within predefined bounds. This design choice ensures that the conclusions drawn from mutation analysis are statistically reliable and methodologically sound.

In conclusion, QCRMut advances the state of the art by bringing mutation testing closer to the quantum paradigm and by providing a flexible approach to the generation and evaluation of quantum mutants, the support for both random and exhaustive strategies under practical constraints, and the integration of an automated testing framework. In addition, we have identified several promising directions for future research aimed at further improving the robustness, scalability of quantum software testing.

\newpage
\subsection{Future work}

Moving forward, we plan to extend this work by addressing several of the limitations and open questions identified in the threats to validity. In particular, we aim to provide additional case studies demonstrating how QCRMut can be tailored to meet diverse testing requirements across different classes of quantum algorithms, thereby strengthening its internal and external validity. This includes exploring the impact of varying circuit sizes, especially with respect to the number of qubits, in order to better understand how the proposed design decisions scale beyond the setting considered in this study.

Another important direction for future research concerns the construct validity of the framework. Several design parameters, such as the chosen representativity threshold, the discretisation of parameter values, and the reduction strategy used in exhaustive mutant generation, warrant a more systematic analysis. Investigating how these parameters influence mutant diversity, coverage, and testing effectiveness would provide deeper insight into the trade-offs between computational cost and testing accuracy.

From a tooling perspective, we intend to explore the adaptation of
QCRMut to intermediate representations such as QASM, which would
facilitate its integration with other quantum software development
kits (SDKs) beyond Qiskit. This would also enable more direct
comparisons with existing mutation testing tools and contribute to
improved external validity. More broadly, we aim to further bridge the
gap between classical and quantum testing paradigms. In particular, we
plan to develop a dedicated abstraction layer for the automated
testing framework, decoupled from the boundaries of QCRMut, so that it
can be fully instantiated and reused independently of the mutant
generation process. This would support more modular, extensible, and
trustworthy quantum software testing, and facilitate its adoption
across a wider range of quantum development workflows.

\appendix
\section{Mutation Experiment - Figures}
\label{A1:AppendixMutFig}

\begin{figure}[H]
    \centering
    \resizebox{0.9\columnwidth}{!}{%
    \includegraphics[width=\textwidth]{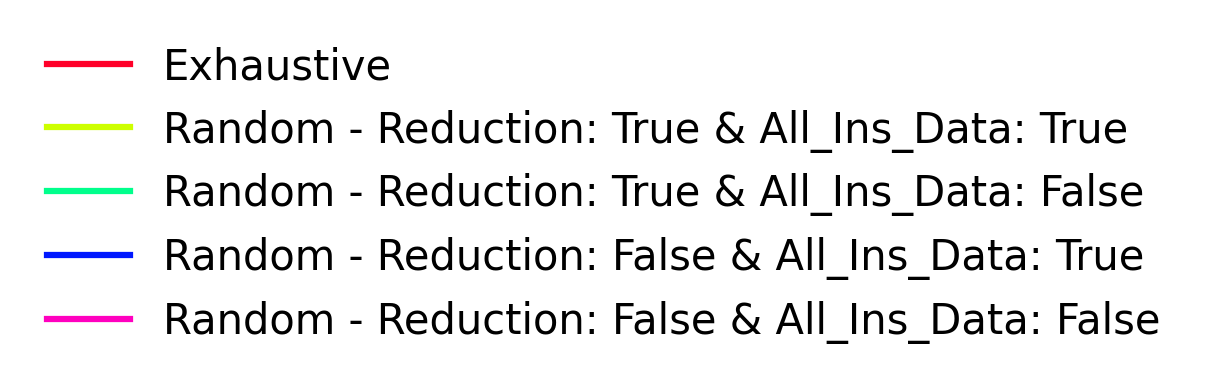}%
    }
    \caption{Legend for the figures presented in Appendix~\ref{A1:AppendixMutFig}.}
\end{figure}

\begin{figure}[H]
    \centering
    \resizebox{0.9\columnwidth}{!}{%
    \includegraphics[width=\textwidth]{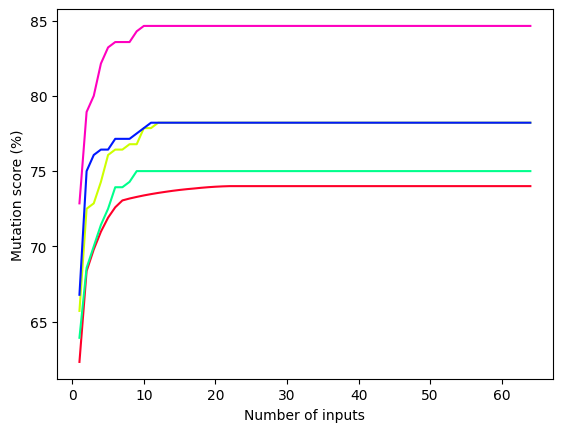}%
    }
    \caption{Mutation score evolution; BV.}
\end{figure}

\begin{figure}[H]
    \centering
    \resizebox{0.9\columnwidth}{!}{%
    \includegraphics[width=\textwidth]{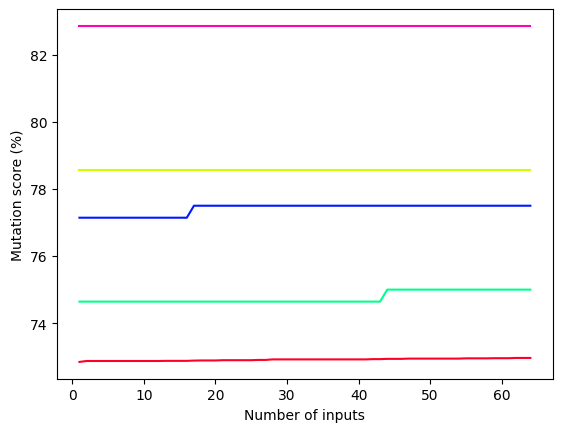}%
    }
    \caption{Mutation score evolution; DJ.}
\end{figure}

\begin{figure}[H]
    \centering
    \resizebox{0.9\columnwidth}{!}{%
    \includegraphics[width=\textwidth]{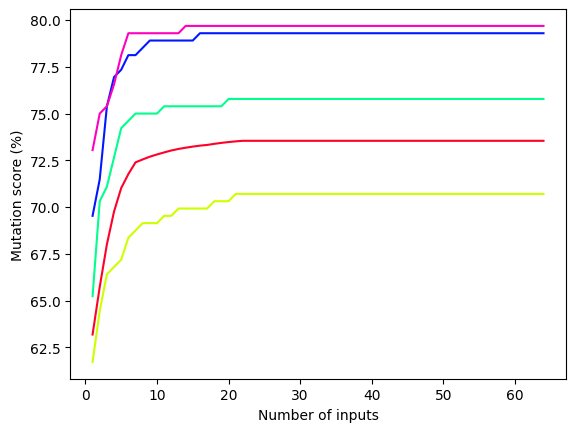}%
    }
    \caption{Mutation score evolution; ghz.}
\end{figure}

\begin{figure}[H]
    \centering
    \resizebox{0.9\columnwidth}{!}{%
    \includegraphics[width=\textwidth]{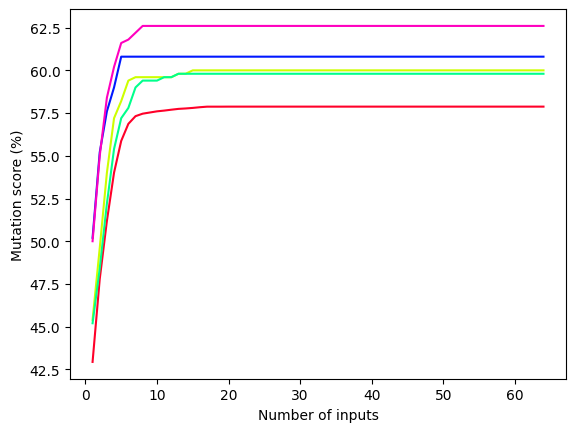}%
    }
    \caption{Mutation score evolution; IQFT.}
\end{figure}

\begin{figure}[H]
    \centering
    \resizebox{0.9\columnwidth}{!}{%
    \includegraphics[width=\textwidth]{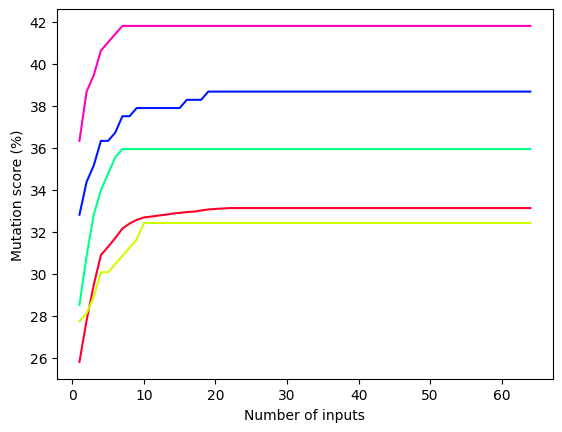}%
    }
    \caption{Mutation score evolution; LittleBV.}
\end{figure}

\begin{figure}[H]
    \centering
    \resizebox{0.9\columnwidth}{!}{%
    \includegraphics[width=\textwidth]{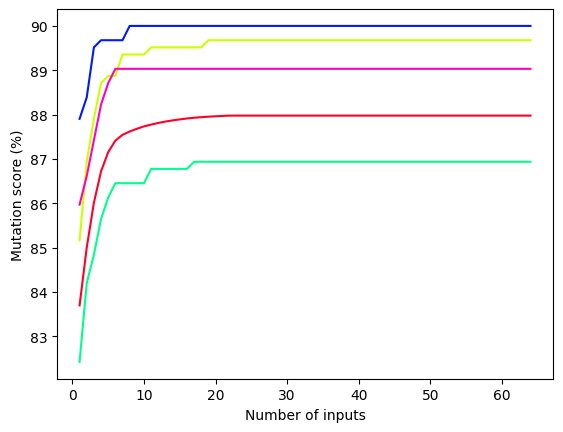}%
    }
    \caption{Mutation score evolution; qftentangled.}
\end{figure}

\begin{figure}[H]
    \centering
    \resizebox{0.9\columnwidth}{!}{%
    \includegraphics[width=\textwidth]{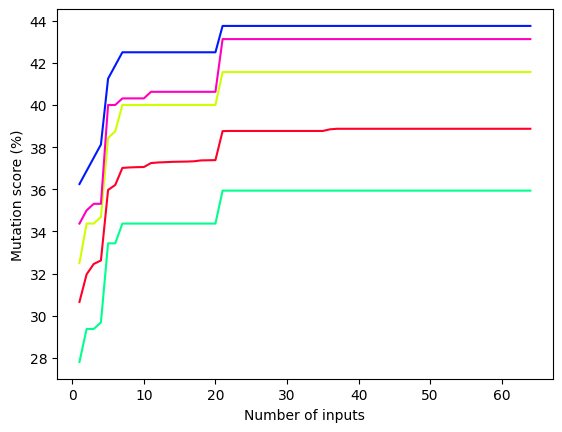}%
    }
    \caption{Mutation score evolution; CE.}
\end{figure}

\begin{figure}[H]
    \centering
    \resizebox{0.9\columnwidth}{!}{%
    \includegraphics[width=\textwidth]{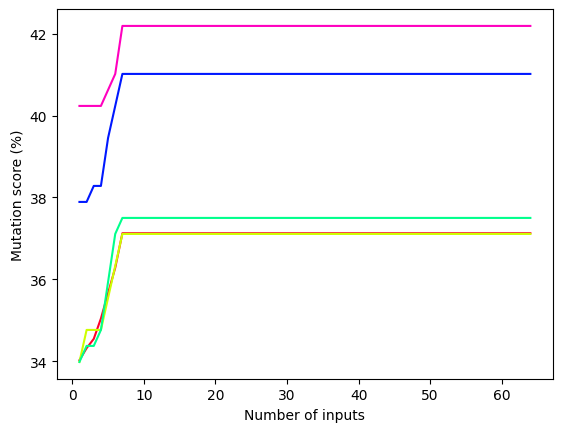}%
    }
    \caption{Mutation score evolution; graphstate.}
\end{figure}

\begin{figure}[H]
    \centering
    \resizebox{0.9\columnwidth}{!}{%
    \includegraphics[width=\textwidth]{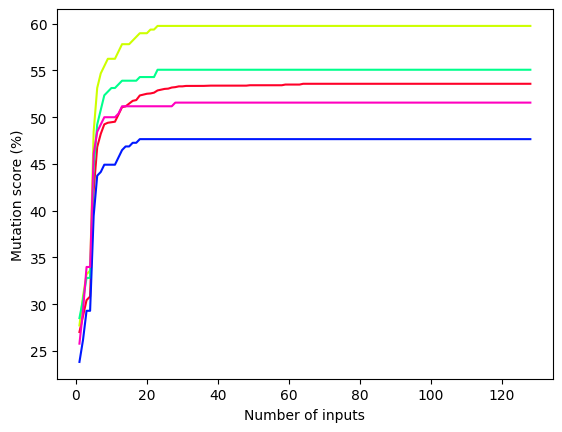}%
    }
    \caption{Mutation score evolution; QRAM.}
\end{figure}

\begin{figure}[H]
    \centering
    \resizebox{0.9\columnwidth}{!}{%
    \includegraphics[width=\textwidth]{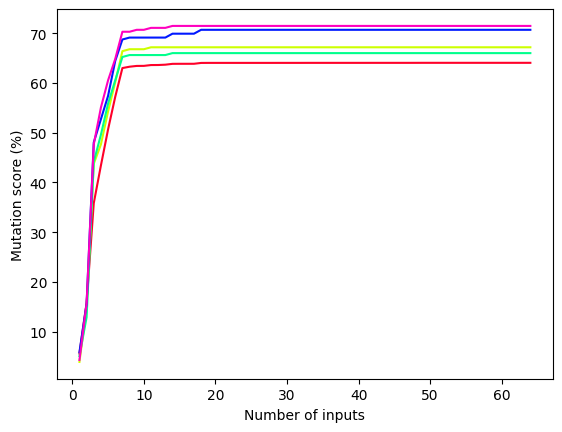}%
    }
    \caption{Mutation score evolution; Simon.}
\end{figure}

\begin{figure}[H]
    \centering
    \resizebox{0.9\columnwidth}{!}{%
    \includegraphics[width=\textwidth]{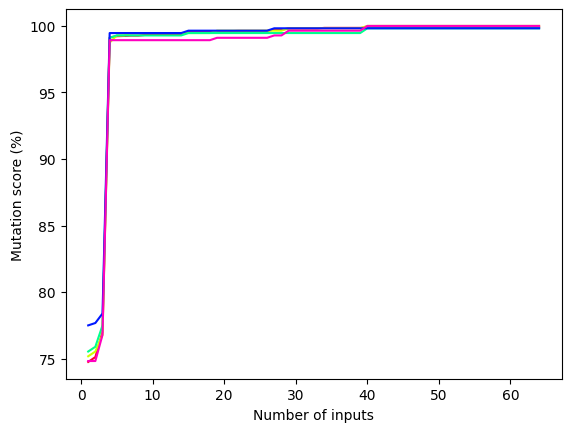}%
    }
    \caption{Mutation score evolution; qpeexact.}
\end{figure}

\begin{figure}[H]
    \centering
    \resizebox{0.9\columnwidth}{!}{%
    \includegraphics[width=\textwidth]{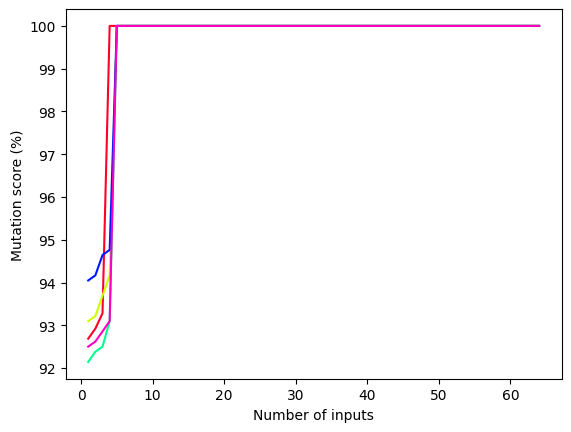}%
    }
    \caption{Mutation score evolution; ae.}
\end{figure}

\begin{figure}[H]
    \centering
    \resizebox{0.9\columnwidth}{!}{%
    \includegraphics[width=\textwidth]{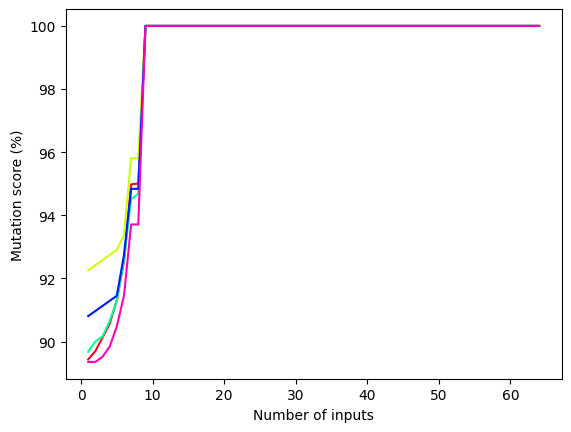}%
    }
    \caption{Mutation score evolution; qaoa.}
\end{figure}

\begin{figure}[H]
    \centering
    \resizebox{0.9\columnwidth}{!}{%
    \includegraphics[width=\textwidth]{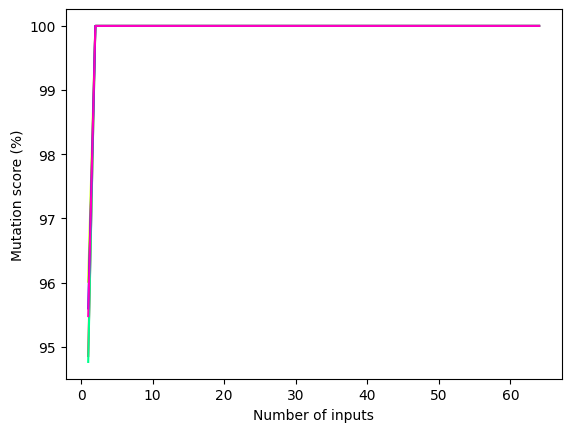}%
    }
    \caption{Mutation score evolution; realamprandom.}
\end{figure}

\begin{figure}[H]
    \centering
    \resizebox{0.9\columnwidth}{!}{%
    \includegraphics[width=\textwidth]{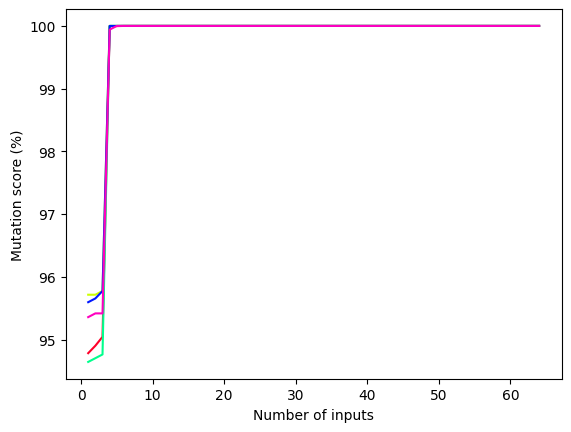}%
    }
    \caption{Mutation score evolution; su2random.}
\end{figure}

\begin{figure}[H]
    \centering
    \resizebox{0.9\columnwidth}{!}{%
    \includegraphics[width=\textwidth]{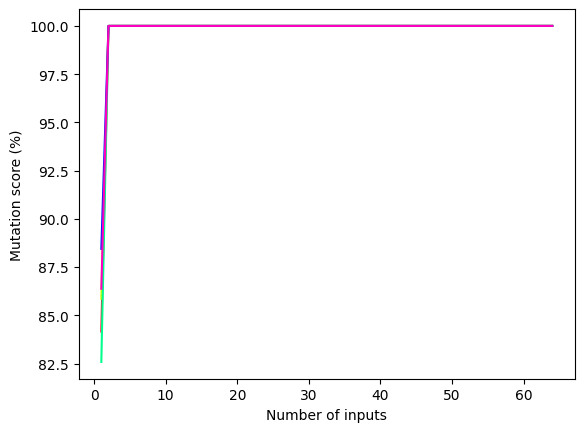}%
    }
    \caption{Mutation score evolution; vqe.}
\end{figure}

\begin{figure}[H]
    \centering
    \resizebox{0.9\columnwidth}{!}{%
    \includegraphics[width=\textwidth]{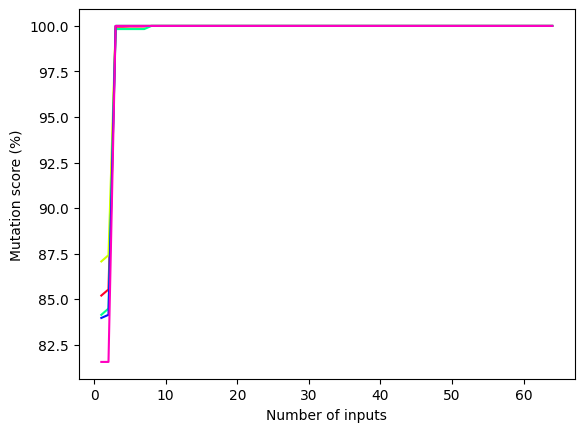}%
    }
    \caption{Mutation score evolution; qpeinexact.}
\end{figure}

\begin{figure}[H]
    \centering
    \resizebox{0.9\columnwidth}{!}{%
    \includegraphics[width=\textwidth]{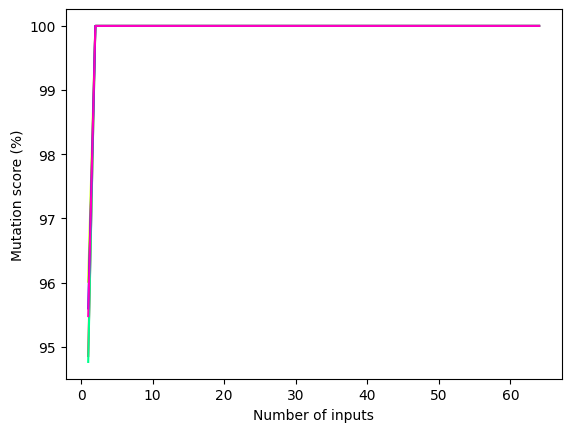}%
    }
    \caption{Mutation score evolution; twolocalrandom.}
\end{figure}

\begin{figure}[H]
    \centering
    \resizebox{0.9\columnwidth}{!}{%
    \includegraphics[width=\textwidth]{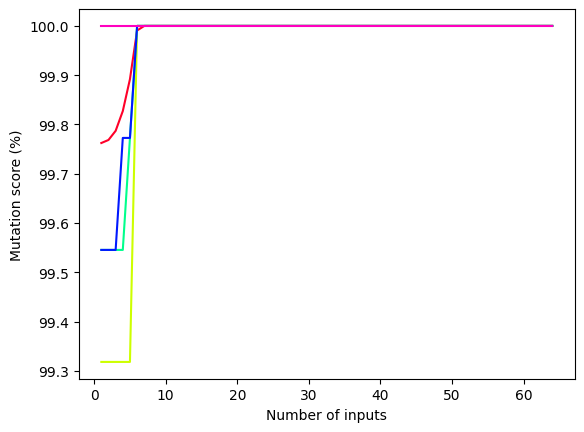}%
    }
    \caption{Mutation score evolution; wstate.}
\end{figure}

\begin{figure}[H]
    \centering
    \resizebox{0.9\columnwidth}{!}{%
    \includegraphics[width=\textwidth]{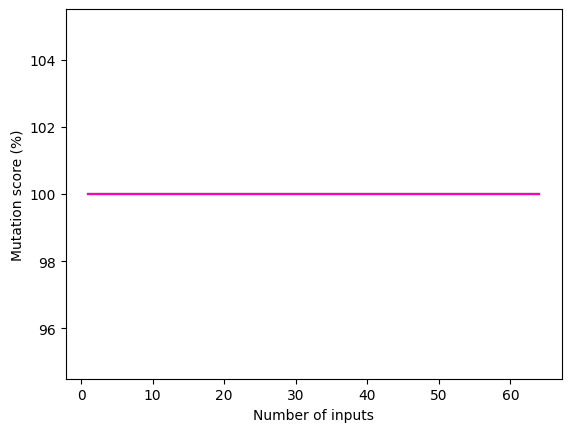}%
    }
    \caption{Mutation score evolution; qft and portfoliovqe.}
\end{figure}
\section{Seed Experiment - Figures}
\label{A2:AppendixSeedFig}

\begin{figure}[H]
    \centering
    \resizebox{\columnwidth}{!}{%
    \includegraphics[width=\textwidth]{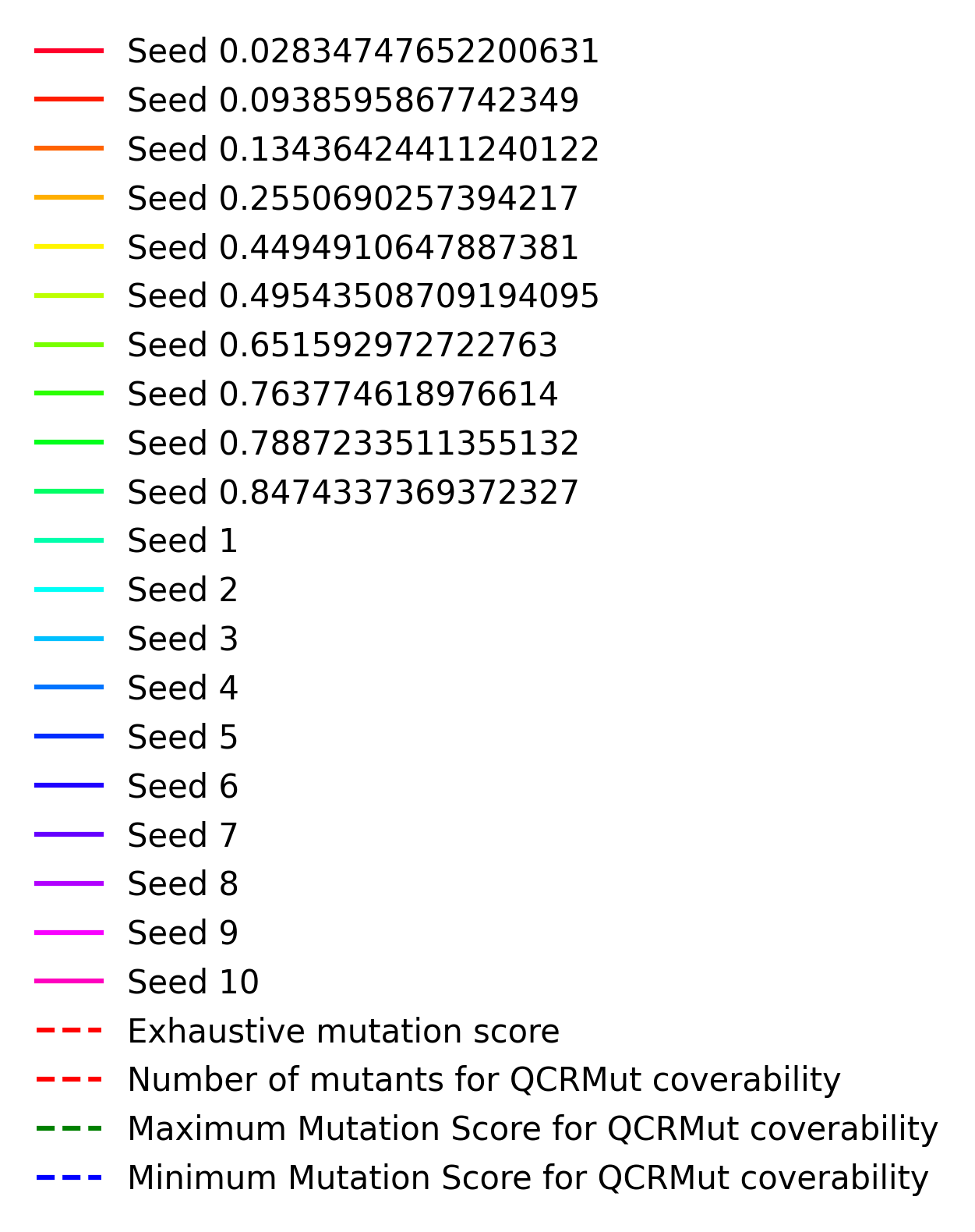}%
    }
    \caption{Legend for the figures presented in Appendix~\ref{A2:AppendixSeedFig}.}
\end{figure}

\begin{figure}[H]
    \centering
    \resizebox{0.9\columnwidth}{!}{%
    \includegraphics[width=\textwidth]{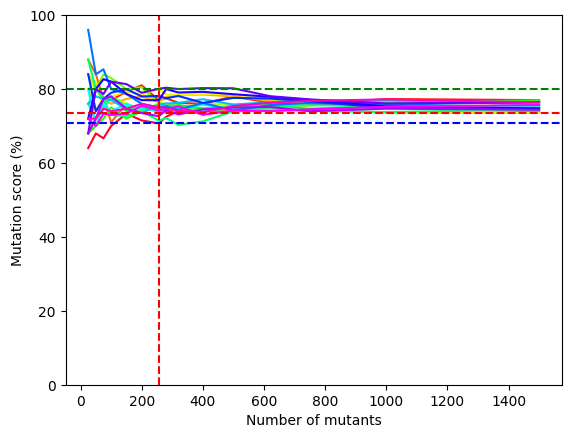}%
    }
    \caption{Mutation score evolution; ghz.}
\end{figure}

\begin{figure}[H]
    \centering
    \resizebox{0.9\columnwidth}{!}{%
    \includegraphics[width=\textwidth]{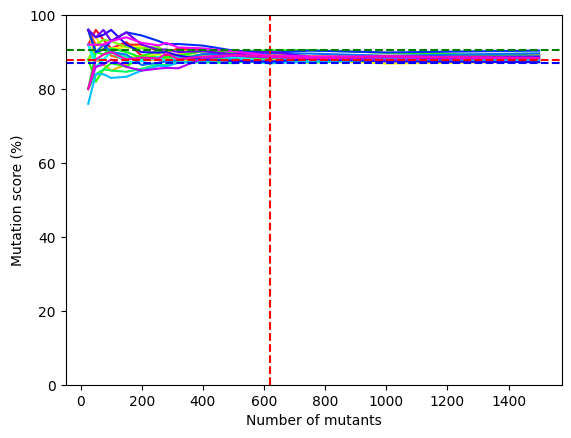}%
    }
    \caption{Mutation score evolution; qftentangled.}
\end{figure}

\begin{figure}[H]
    \centering
    \resizebox{0.9\columnwidth}{!}{%
    \includegraphics[width=\textwidth]{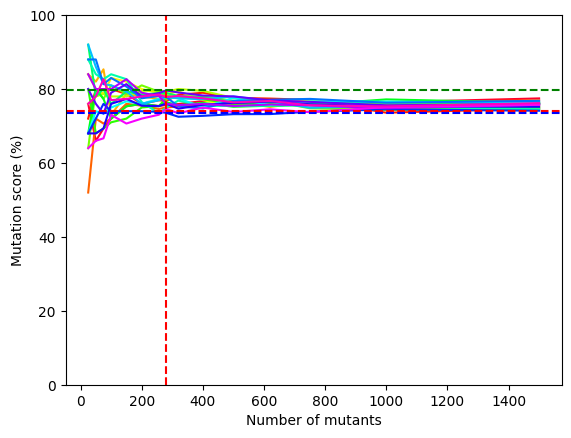}%
    }
    \caption{Mutation score evolution; BV.}
\end{figure}

\begin{figure}[H]
    \centering
    \resizebox{0.9\columnwidth}{!}{%
    \includegraphics[width=\textwidth]{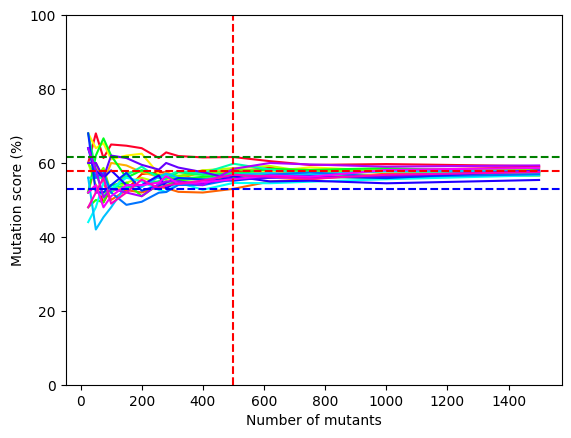}%
    }
    \caption{Mutation score evolution; IQFT.}
\end{figure}

\begin{figure}[H]
    \centering
    \resizebox{0.9\columnwidth}{!}{%
    \includegraphics[width=\textwidth]{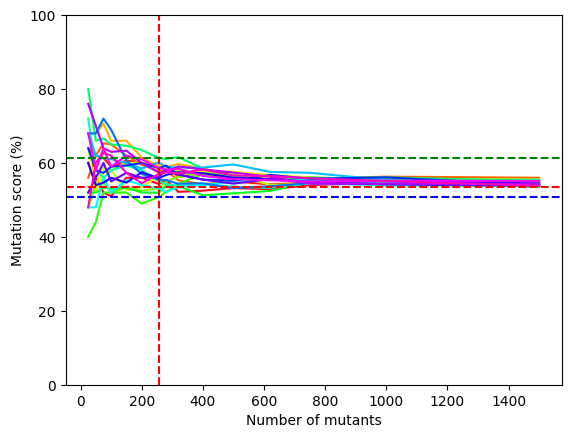}%
    }
    \caption{Mutation score evolution; QRAM.}
\end{figure}

\begin{figure}[H]
    \centering
    \resizebox{0.9\columnwidth}{!}{%
    \includegraphics[width=\textwidth]{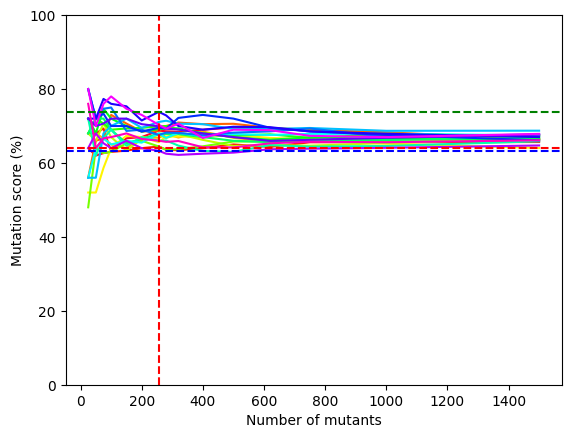}%
    }
    \caption{Mutation score evolution; Simon.}
\end{figure}

\begin{figure}[H]
    \centering
    \resizebox{0.9\columnwidth}{!}{%
    \includegraphics[width=\textwidth]{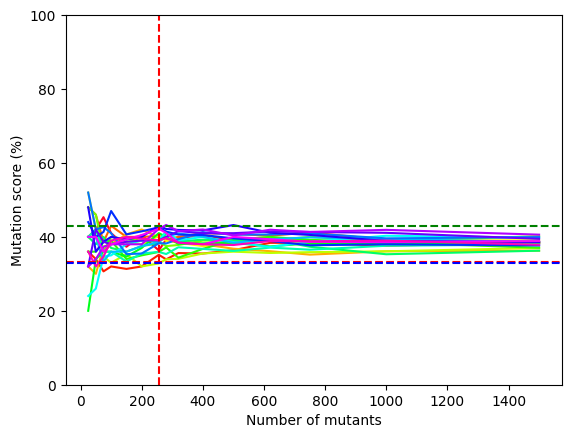}%
    }
    \caption{Mutation score evolution; LittleBV.}
\end{figure}

\begin{figure}[H]
    \centering
    \resizebox{0.9\columnwidth}{!}{%
    \includegraphics[width=\textwidth]{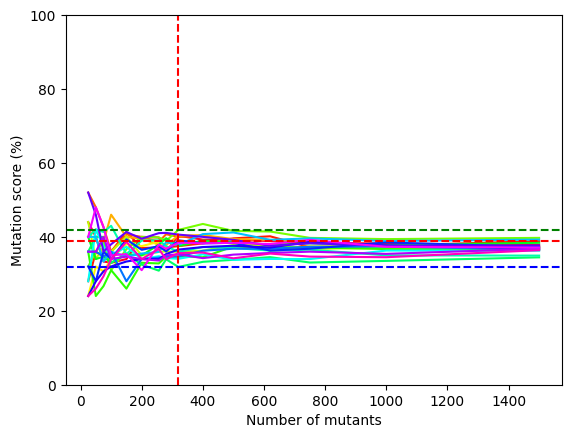}%
    }
    \caption{Mutation score evolution; CE.}
\end{figure}

\begin{figure}[H]
    \centering
    \resizebox{0.9\columnwidth}{!}{%
    \includegraphics[width=\textwidth]{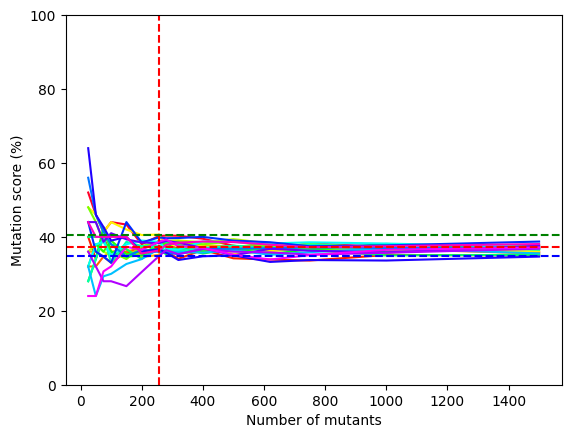}%
    }
    \caption{Mutation score evolution; graphstate.}
\end{figure}

\bibliographystyle{IEEEtran}
\bibliography{bibliography.bib}

\end{document}